\DocumentMetadata{}
\documentclass[sigconf]{acmart}
\usepackage{balance}       %
\usepackage{graphics}      %
\usepackage{hyperref}
\usepackage{color}
\usepackage{booktabs}
\usepackage{textcomp}
\usepackage{subcaption}
\usepackage{enumerate}
\usepackage{xcolor}
\usepackage{lipsum}%
\usepackage{makecell}
\usepackage{multicol}
\usepackage{multirow}
\usepackage{array}
\usepackage{verbatimbox}
\usepackage{enumitem}
\usepackage{amsmath}
\usepackage{stfloats}
\usepackage{graphicx}
\usepackage{amsthm}
\usepackage{listings}
\usepackage{caption} 
\usepackage{xspace}
\usepackage[linesnumbered]{algorithm2e}
\usepackage{algpseudocode}
\usepackage{tabularx}
\usepackage{arydshln}
\usepackage[bottom]{footmisc}
\usepackage{stackengine}
\usepackage{placeins}
\usepackage{CJKutf8}
\usepackage{float} 
\usepackage{xcolor}
\usepackage[normalem]{ulem}

\definecolor{myYellow}{RGB}{197, 148, 3}
\definecolor{myBlue}{RGB}{25, 114, 222}
\definecolor{myRed}{RGB}{215, 85, 85}
\definecolor{myGreen}{RGB}{56, 168, 34}

\newcommand{\textyellow}[1]{\textcolor{myYellow}{#1}}
\newcommand{\textblue}[1]{\textcolor{myBlue}{#1}}
\newcommand{\textred}[1]{\textcolor{myRed}{#1}}
\newcommand{\textgreen}[1]{\textcolor{myGreen}{#1}}

\newcommand*{\eg}{\textit{e.g.},\xspace}

\newcolumntype{L}[1]{>{\raggedright\let\newline\\\arraybackslash\hspace{0pt}}m{#1}}
\newcolumntype{C}[1]{>{\centering\let\newline\\\arraybackslash\hspace{0pt}}m{#1}}
\newcolumntype{R}[1]{>{\raggedleft\let\newline\\\arraybackslash\hspace{0pt}}m{#1}}

\makeatletter
\def\thickhline{%
  \noalign{\ifnum0=`}\fi\hrule \@height \thickarrayrulewidth \futurelet
   \reserved@a\@xthickhline}
\def\@xthickhline{\ifx\reserved@a\thickhline
               \vskip\doublerulesep
               \vskip-\thickarrayrulewidth
             \fi
      \ifnum0=`{\fi}}
\makeatother

\makeatletter
\def\thickhlinespace{%
  \addlinespace[1ex]
  \noalign{\ifnum0=`}\fi\hrule \@height \thickarrayrulewidth \futurelet
   \reserved@a\@xthickhline
   \addlinespace[1ex]
   }
\def\@xthickhlinespace{\ifx\reserved@a\thickhline
               \vskip\doublerulesep
               \vskip-\thickarrayrulewidth
             \fi
      \ifnum0=`{\fi}}
\makeatother

\newlength{\thickarrayrulewidth}
\newlength\Origarrayrulewidth

\algnewcommand{\IfThenElse}[3]{%
  \State \algorithmicif\ #1\ \algorithmicthen\ #2\ \algorithmicelse\ #3}

\definecolor{DarkGreen}{HTML}{5DAC81}

\newcommand\yuling[1]{\textcolor{black}{#1}}

\definecolor{downredcolor}{HTML}{e31a1c}
\definecolor{upgreencolor}{HTML}{33a02c}

 \newcommand\review[1]{\textcolor{black}{#1}}
\newcommand\minorreview[1]{\textcolor{black}{#1}}

\usepackage{tikz}
\usepackage{xcolor}
\usepackage{xparse}

\definecolor{plotdata}{HTML}{54A570}       %
\definecolor{plotmean}{HTML}{0E7A33}       %
\definecolor{plotmeanlabel}{HTML}{994D00}  %
\definecolor{plotticks}{HTML}{666666}      %
\newlength{\dotplotheight}
\newlength{\dotplotwidth}
\pgfmathsetlengthmacro{\plotwidthnum}{\dotplotwidth}
\pgfmathsetlengthmacro{\plotheightnum}{\dotplotheight}

\pgfmathsetmacro{\tickh}{0.2}    
\pgfmathsetmacro{\meanh}{0.5}     
\pgfmathsetmacro{\pointsize}{0.12}

\ExplSyntaxOn
\keys_define:nn { dotplot }
 {
  min  .tl_set:N  = \l_dotplot_min_tl,
  min  .initial:n = {0},
  max  .tl_set:N  = \l_dotplot_max_tl,
  max  .initial:n = {10},
  step .tl_set:N  = \l_dotplot_step_tl,
  step .initial:n = {1},
 }
\ExplSyntaxOff

\ExplSyntaxOn
\NewDocumentCommand{\dotplot}{ O{} m }
 {
  \group_begin:
  \keys_set:nn { dotplot } { #1 }

  \pgfmathsetmacro{\dotplotSum}{0}
  \pgfmathsetmacro{\dotplotCount}{0}
  \foreach \val in {#2} {
      \pgfmathparse{\dotplotSum + \val} \global\let\dotplotSum\pgfmathresult
      \pgfmathparse{\dotplotCount + 1} \global\let\dotplotCount\pgfmathresult
  }
  \pgfmathsetmacro{\mean}{ifthenelse(\dotplotCount>0,\dotplotSum/\dotplotCount,0)}

  \pgfmathsetmacro{\numticks}{int((\l_dotplot_max_tl - \l_dotplot_min_tl)/\l_dotplot_step_tl)}

\begin{tikzpicture}[x=\plotwidthnum/(\l_dotplot_max_tl - \l_dotplot_min_tl), y=\plotheightnum, baseline=-0.25\dotplotheight]
    \draw[plotticks] (\l_dotplot_min_tl,0) -- (\l_dotplot_max_tl,0);

    \pgfmathsetlengthmacro{\tickheightnum}{\tickh*\dotplotheight}
    \foreach \i in {0,...,\numticks} {
        \pgfmathsetmacro{\x}{\l_dotplot_min_tl + \i*\l_dotplot_step_tl}
        \ifnum\i=0
            \draw[plotticks, thick] (\x,-\tickheightnum) -- (\x,\tickheightnum);
            \node[overlay, font=\tiny, anchor=north, gray] at (\x+0.15,\tickheightnum-1) {\l_dotplot_min_tl};
        \else
            \ifnum\i=\numticks
                \draw[plotticks, thick] (\x,-\tickheightnum) -- (\x,\tickheightnum);
                \node[overlay, font=\tiny, anchor=north, gray] at (\x-0.15,\tickheightnum-1) {\l_dotplot_max_tl};
            \else
                \draw[plotticks] (\x,0) -- (\x,\tickheightnum);
            \fi
        \fi
    }

    \pgfmathsetlengthmacro{\radiusnum}{\pointsize*\dotplotheight}
    \def\jitter{0.15} %
    
    \foreach \val in {#2} {
      \pgfmathsetmacro{\yjitter}{\jitter*rand}
      \fill[plotdata, opacity = 0.6] (\val,\yjitter) circle[radius=\radiusnum];
    }

    \pgfmathsetlengthmacro{\meanheightnum}{\meanh*\dotplotheight}
    \draw[plotmean, line~width=1pt] (\mean,-\meanheightnum) -- (\mean,\meanheightnum);
    \node[overlay, font={\bfseries\tiny}, anchor=west, gray, plotmean] 
    at (\mean-0.1, \tickheightnum+2.2) 
    {\pgfmathprintnumber[fixed, precision=2]{\mean}};
  \end{tikzpicture}

  \group_end:
 }
\ExplSyntaxOff

\ExplSyntaxOn
\NewDocumentCommand{\pieplot}{ O{plotticks} m m }
 {
  \pgfmathsetmacro{\pieangle}{360*(#2)/(#3)}
  
  \begin{tikzpicture}[baseline=-0.8ex]
    \draw[#1] (0,0) circle (0.13cm);
    \fill[#1!60] (0,0) -- (0,0.13cm)
      arc [start~angle=90, end~angle={90-\pieangle}, radius=0.13cm] -- cycle;
  \end{tikzpicture}%
 }
\ExplSyntaxOff

\newcommand{\progressratio}[4][3.0]{%
  \begin{tikzpicture}[baseline=-0.25ex]
    \pgfmathsetmacro{\percentage}{(#2 - #3)/(#4 - #3)}
    \pgfmathsetmacro{\barwidth}{#1}
    \pgfmathsetmacro{\progresswidth}{\percentage * \barwidth}
    
    \fill[color=lightgray!50] (0, 0) rectangle +(\barwidth ex, 1.2ex);
    
    \fill[color=plotdata!80] (0, 0) rectangle +(\progresswidth ex, 1.2ex);

    \node[font=\small, anchor=west, text=black, inner sep=1pt] 
      at (\barwidth ex + 0.3ex, 0.6ex) 
      {#2/#4}; 
  \end{tikzpicture}%
}

\newcommand{\progressratioTotal}[4][3.0]{%
  \begin{tikzpicture}[baseline=-0.25ex]
    \pgfmathsetmacro{\percentage}{(#2 - #3)/(#4 - #3)}
    \pgfmathsetmacro{\barwidth}{#1}
    \pgfmathsetmacro{\progresswidth}{\percentage * \barwidth}
    
    \fill[color=lightgray!50] (0, 0) rectangle +(\barwidth ex, 1.2ex);
    
    \fill[color=plotmean] (0, 0) rectangle +(\progresswidth ex, 1.2ex); 

    \node[font=\small, anchor=west, text=black, inner sep=1pt] 
      at (\barwidth ex + 0.3ex, 0.6ex) 
      {#2/#4}; 
  \end{tikzpicture}%
}

\copyrightyear{2026}
\acmYear{2026}
\setcopyright{cc}
\setcctype{by}
\acmConference[CHI '26]{Proceedings of the 2026 CHI Conference on Human Factors in Computing Systems}{April 13--17, 2026}{Barcelona, Spain}
\acmBooktitle{Proceedings of the 2026 CHI Conference on Human Factors in Computing Systems (CHI '26), April 13--17, 2026, Barcelona, Spain}
\acmPrice{}
\acmDOI{10.1145/3772318.3791946}
\acmISBN{979-8-4007-2278-3/2026/04}

\begin{document}

\title[LLM Doctors as Longitudinal Boundary Companions]{More than Decision Support: Exploring Patients' Longitudinal Usage of Large Language Models in \yuling{Real-World Healthcare-Seeking Journeys}}

\author{Yancheng Cao}
\orcid{0000-0003-3033-8881}
\affiliation{%
  \institution{Columbia University}
  \city{New York}
  \state{New York}
  \country{US}}
\email{yanchengc7@outlook.com}

\author{Yishu Ji}
\orcid{0009-0005-4462-7694}
\affiliation{%
  \institution{Georgia Institute of Technology}
  \city{Atlanta}
  \state{Georgia}
  \country{US}}
\email{yji329@gatech.edu}

\author{Yue Fu}
\orcid{0000-0001-5828-5932}
\affiliation{%
  \institution{University of Washington}
  \city{Seattle}
  \state{Washington}
  \country{US}}
\email{chrisfu@uw.edu}

\author{Sahiti Dharmavaram}
\orcid{0009-0000-0616-4258}
\affiliation{%
  \institution{Columbia University}
  \city{New York}
  \state{New York}
  \country{US}}
\email{sd3976@columbia.edu}

\author{Meghan Turchioe}
\orcid{0000-0002-6264-6320}
\affiliation{%
  \institution{Columbia University}
  \city{New York}
  \state{New York}
  \country{US}}
\email{mr3554@cumc.columbia.edu}

\author{Natalie C Benda}
\orcid{0000-0002-3256-0243}
\affiliation{%
  \institution{Columbia University}
  \city{New York}
  \state{New York}
  \country{US}}
\email{nb3115@cumc.columbia.edu}

\author{Lena Mamykina}
\orcid{0000-0001-5203-274X}
\affiliation{%
  \institution{Columbia University}
  \city{New York}
  \state{New York}
  \country{US}}
\email{om2196@cumc.columbia.edu}

\author{Yuling Sun}
\authornote{Mark corresponding authors.}
\orcid{0000-0003-1726-5913}
\affiliation{%
  \institution{University of Michigan}
  \city{Ann Arbor}
  \state{Michigan}
  \country{US}}
\email{yulingsu@umich.edu}

\author{Xuhai Xu}
\authornotemark[1]
\orcid{0000-0001-5930-3899}
\affiliation{%
  \institution{Columbia University}
  \city{New York}
  \state{New York}
  \country{US}}
\email{xx2489@columbia.edu}

\renewcommand{\shortauthors}{Cao et al.}
\renewcommand{\shorttitle}{More than Decision Support}

\begin{abstract}
Large language models (LLMs) have been increasingly adopted to support patients' healthcare-seeking in recent years. While prior patient-centered studies have examined the capabilities and experience of LLM-based tools in specific health-related tasks such as information-seeking, diagnosis, or decision-supporting, the inherently longitudinal nature of healthcare in real-world practice has been underexplored. This paper presents a four-week diary study with 25 patients to examine LLMs' roles across healthcare-seeking trajectories.
Our analysis reveals that patients integrate LLMs not just as simple decision-support tools, but as dynamic companions that scaffold their journey across behavioral, informational, emotional, and cognitive levels.
Meanwhile, patients actively assign diverse socio-technical meanings to LLMs, altering the traditional dynamics of agency, trust, and power in patient-provider relationships. 
Drawing from these findings, we conceptualize future LLMs as a longitudinal boundary companion that continuously mediates between patients and clinicians throughout longitudinal healthcare-seeking trajectories.

\end{abstract}


\maketitle

\begin{figure*}
\centering 
\includegraphics[width=\linewidth]{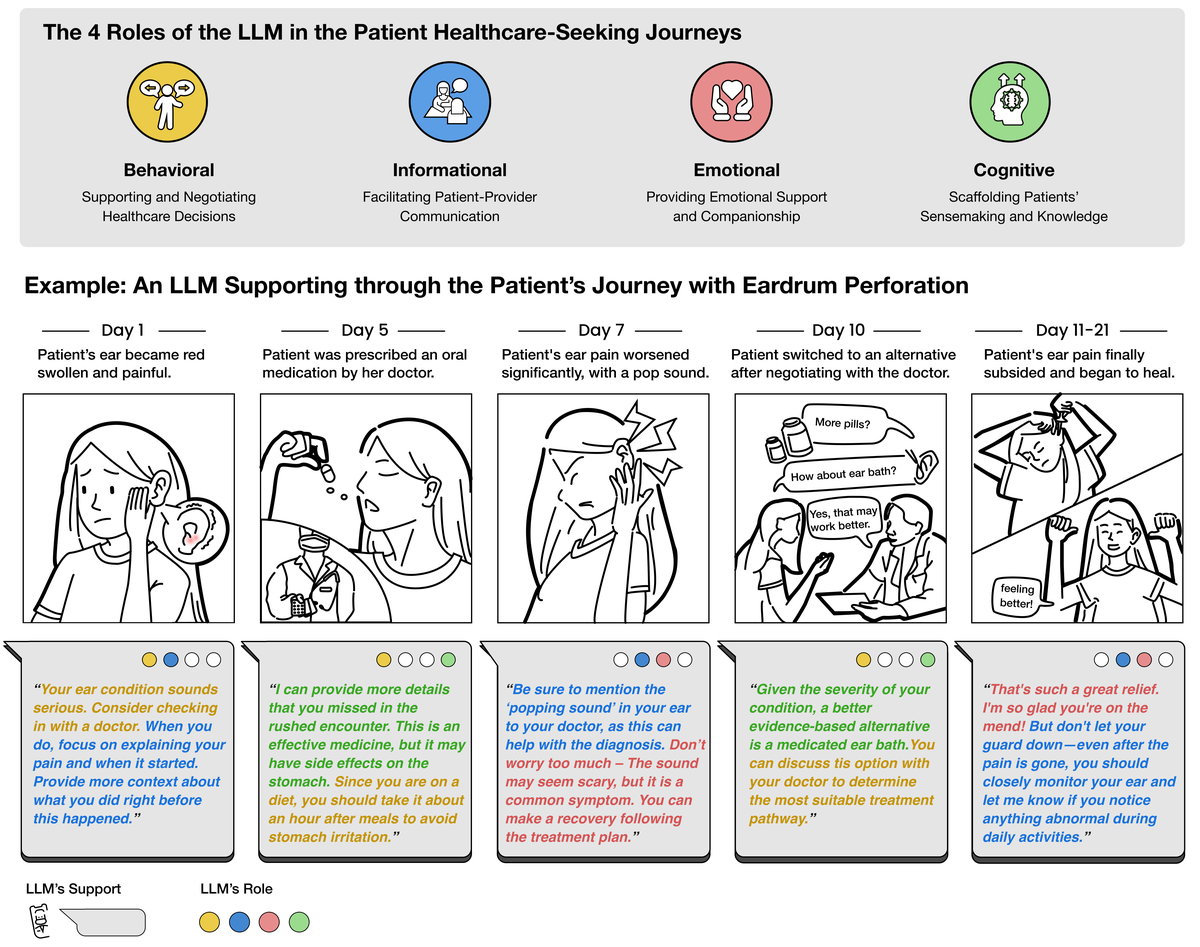}
\caption{\minorreview{Overview of the LLM's four dynamic roles integrated into patients' healthcare-seeking journeys: (1) Behavioral—supporting and negotiating healthcare decisions; (2) Informational—facilitating patient–provider communication; (3) Emotional—providing emotional support and companionship; and (4) Cognitive—scaffolding patients’ sensemaking and knowledge in understanding professional healthcare information.
These roles accompany patients across multiple stages of their journeys, dynamically shifting to meet patients' evolving needs. A case of a patient with eardrum perforation illustrates how LLMs enact these roles at different time points along the healthcare-seeking journey.}
}
\label{fig:teaser}
\Description{
Overview of the four dynamic roles adopted by LLMs and integrated into patients’ healthcare-seeking journeys. On the top, a title reads "The 4 Roles of LLMs in the Patient Healthcare-Seeking Journey." Below are four icons representing each role: Behavioral (yellow icon): "Supporting and negotiating healthcare decisions."; Informational (blue icon): "Facilitating patient–provider communication."; Emotional (red icon): "Providing Emotional Support and Companionship (green icon): "Scaffolding patients’ sensemaking and knowledge." Bottom part, a five-panel comic strip provides an example: "An LLM Supporting the Patient’s Journey with Eardrum Perforation." Each panel shows the patient’s condition alongside a corresponding response from the LLM, color-coded to the roles described above. Day 1: A woman holds her painful, swollen ear. The LLM’s response is Behavioral (yellow) and Informational (blue), analyzing her condition and highlighting key points for discussion with her physician. Day 5: The patient is prescribed oral medication by her doctor. The LLM’s response is Cognitive (green) and Behavioral (yellow), providing medication details the patient missed in a rushed consultation and offering guidance on proper usage. Day 7: The woman clutches her ear in worsened pain after hearing a "pop sound." The LLM’s response is Informational (blue) and Emotional (red), advising her to inform her physician and offering reassurance during this critical moment. Day 10: The patient discusses alternative treatments with her doctor. The LLM’s response is Cognitive (green) and Behavioral (yellow), giving supplementary information and suggesting she raise this option in her medical discussion. Day 11–21: The woman is smiling and feeling better. The LLM’s response is Informational (blue) and Emotional (red), providing companionship and reminding her to continue monitoring her condition and follow up as needed.
}
\end{figure*}

\begin{CJK*}{UTF8}{gbsn}

\section{Introduction}
\label{sec:introduction}

\yuling{
The practical healthcare-seeking process is not a single event but a dynamic, multi-stage trajectory, covering symptom appraisal, information seeking, clinical encounter, post-consultation sensemaking, treatment planning and implementation, recurrence monitoring, and follow-up cycles \cite{andersen1995revisiting}. Throughout this journey, patients are often placed in a vulnerable position \cite{balint1955doctor, moos1977crisis}. They have to continually cope with the uncertainty of their symptoms while also navigating persistent challenges such as limited information, the structural complexity of healthcare systems, and considerable emotional stress \cite{guadagnoli1998patient}. 
These challenges are further amplified in resource-constrained healthcare settings, where limited healthcare resources, information asymmetries, and insufficient primary or long-term care support remain widespread \cite{liang2021patient, yan2021patient,chen2021ten}. As a result, patients are increasingly struggling to access timely, adequate, and personalized healthcare support.}

\yuling{To bridge this gap, patients have increasingly turned to digital artificial intelligence (AI) technologies, ranging from search engines for information retrieval~\cite{luo2008medsearch, bernstam2005instruments} to specialized symptom checkers~\cite{semigran2015evaluation, meyer2020patient} and largely-scripted medical chatbots~\cite{rosruen2018chatbot, xu2021chatbot}.
While these tools democratized access to additional healthcare support, they were often limited by rigid interaction patterns and a lack of contextual understanding, struggling to address the emotional and shifting needs of a patient's dynamic and longitudinal healthcare-seeking journey.}

In recent years, Generative AI, in particular large language models (LLMs), has enabled the boom of conversational AI applications in the healthcare domain~\cite{nori2023capabilities, sarraju2023appropriateness}.
\yuling{
In particular,  with the powerful natural language understanding and generation capabilities \cite{singhal2025toward, goh2025gpt}, LLMs can not only answer medical questions but also help patients articulate symptoms \cite{yang2024talk2care,subramanian2024enhancing}, interpret clinical advice \cite{amin2023artificial,aydin2024large}, plan next steps \cite{lv2024leveraging,wu2024mindshift}, and provide emotional support \cite{jung2025ve,song2025typing} through sustained conversational engagement. Consequently, LLMs have the potential to embed themselves continuously across the entire healthcare-seeking journey, from the earliest stages of symptom appraisal to follow-up and long-term management, forming a new paradigm of collaborative, trajectory-wide healthcare support that fundamentally reshapes existing patient journeys.
}

Nevertheless, as a high-stakes domain, integrating LLM technologies in healthcare, while ensuring their usability, safety, and ethical deployment, remains a complex and challenging endeavor \cite{busch2025current}. \yuling{This requires a systematic and critical examination of LLMs' roles across the full healthcare-seeking journey, as well as their deeper impact on patients’ experiences and the patient–provider relationship.
As a response, existing HCI has begun to explore the practical adoption, use, and impact of LLM-powered healthcare applications on patients' healthcare-seeking process, such as patients' adoption and experience of using LLMs to support their medical decision-making~\cite{lee2025understanding,foo2025benefits}, as well as the factors influencing the adoption and final decisions \cite{kim2024much, detjen2025trusted}. However, existing works have mainly focused on single-interaction or short-term scenarios~\cite{chen2025patient,van2025selective,jung2025ve}, neglecting the multi-stage longitudinal nature of real-world healthcare-seeking journeys.}
This may produce partial or even misleading understandings of how LLMs actually function and impact patients' healthcare-seeking practices in long-term healthcare contexts.
Moreover, existing studies tend to focus on LLMs' roles in specific health-related tasks, such as information seeking \cite{carl2025evaluating}, diagnosis \cite{wang2024beyond}, and decision support~\cite{aydin2025navigating}, under-exploring how patients, across their complex medical journeys, actively ascribe dynamic roles and meanings to LLMs.

\yuling{Our study aims to fill these gaps through exploring the dynamic, evolving roles and situated significance that LLMs play in patients' long-term healthcare-seeking journeys.
We particularly ask the following two research questions: (1) \textbf{What role do LLMs play throughout patients' entire healthcare-seeking journey}, and (2) \textbf{How does the interaction with LLMs (re)shape their healthcare-seeking journeys, as well as the broader patient-provider relationship?} With the growing development, deployment, and adoption of LLM-powered healthcare applications, understanding these questions can help researchers and developers better shed light on how LLMs can be meaningfully designed and integrated into patients’ healthcare-seeking journey in ways that enhance user experience and maximize their potential value.}

We address these research questions through an IRB-approved, 4-week longitudinal diary study with 25 patients undergoing various health conditions\footnote{It is noteworthy that this work particularly focuses on patients' perspectives. LLM-empowered systems that support clinicians and other important stakeholders are also interesting and worth exploration, but they are beyond the scope of this paper.}. 
We recruited participants to observe how they used LLMs in their healthcare-seeking procedures\footnote{To minimize safety and ethical risks, we provided an educational introduction to all participants, stressing that these LLM tools are supplementary and not a replacement for professional medical care.}, and asked them to document their daily usage experiences and reflections in diaries, followed by in-depth interview to deeply capture their experience, perceptions, and meaning-making processes.

Our analysis reveals that LLMs are not used as simple decision-support tools, but are integrated as dynamic companions across patients' long-term healthcare-seeking journeys.
We show how they scaffold patient practices at four key levels across the treatment stages: behavioral (negotiating decisions with human clinicians), informational (improving patient-provider communication), emotional (providing companionship), and cognitive (scaffolding jargon sensemaking and medical knowledge).
Further, this deep integration empowers patients, reconfiguring traditional dynamics of agency, trust, and power between them and human clinicians. Based on these findings, we conceptualize these LLM technologies as a longitudinal boundary companion—a socio-technical partner that mediates between the clinic and daily life, expert knowledge and patient experience, and cognitive needs and emotional support.
This reframing leads us to propose design implications focused on relational continuity, which involves designing LLM systems to build and maintain a relationship over a patient's entire healthcare journey, and responsible empowerment, which supports patient agency while simultaneously mitigating the risks of misinformation.
Meanwhile, we highlight the potential ethical risks associated with such a broader adoption of LLM roles in patients' healthcare-seeking journeys.

We contribute to the HCI community by extending current understandings of LLM-powered applications in healthcare, foregrounding patients' longitudinal engagements with LLMs across their healthcare-seeking journeys. Specifically:
\begin{itemize}
\item We provide rich empirical evidence from a four-week diary study that details the multifaceted and dynamic roles LLMs play across patients' longitudinal healthcare-seeking journeys, spanning behavioral, informational, emotional, and cognitive dimensions.
\item We introduce the concept of the LLM as a longitudinal boundary companion and analyze how it reconfigures patient agency, trust, and power, with its ethical implications. This provides a new theoretical lens for understanding the relational and socio-technical impact of LLMs in healthcare.
\item We articulate a set of concrete design implications that move beyond accuracy and information tools, focusing on relational continuity, responsible empowerment, and scaffolding health literacy in future LLM-powered health applications.
\end{itemize}

\section{Related Work}
\label{sec:related_work}

\yuling{Our study is situated within the broader research landscape of AI-powered healthcare-seeking. In this section, we review existing work along three key lines: patients’ healthcare-seeking journeys (Section \ref{sub:related_work:2.1}), the promise and limitations of LLM-powered healthcare support (Section \ref{sub:related_work:2.2}), and the socio-technical dynamics when adopting LLM-mediated application into real-world healthcare settings (Section \ref{sub:related_work:2.3}).
These three bodies of literature establish the conceptual background for understanding the contribution of our study.}

\subsection{\yuling{Patient Healthcare-Seeking Journey: Longitudinal, Dynamic, and Multi-Stage Trajectory}}
\label{sub:related_work:2.1}

\yuling{Healthcare-seeking is fundamentally a multi-stage and continuously unfolding process \cite{andersen1995revisiting}, typically involving symptom appraisal, information seeking, care navigation and decision-making, the clinical encounter, post-consultation sensemaking and planning, treatment implementation and self-management, and long-term recurrence monitoring and follow-up~\cite{elwyn2012shared,scott2013model}. Each stage is characterized by distinct information needs, emotional states, decision pressures, and resource availability, resulting in constantly shifting challenges for patients throughout their healthcare journey~\cite{de2022integrating,davies2023reporting}. This multi-stage, dynamically evolving nature of patients' needs implies that digital healthcare technologies must be able to adapt responsively across stages in order to achieve sustained, real-world effectiveness~\cite{nahum2016just,glasgow2019adaptive}.}

\yuling{
To achieve such technological design, researchers and designers should systematically investigate patients' evolving needs within real-world healthcare practices and understand how technologies are adopted, interpreted, and integrated across different stages of the healthcare-seeking journey.
This perspective aligns with HCI's longstanding emphasis on studying user needs and technology use within authentic, situated contexts, rather than isolated or decontextualized scenarios \cite{kling2000scientific, suchman1987plans}.
\minorreview{In digital health and personal informatics, many studies have explored patients' needs and technology usage during healthcare-seeking.
For example, conversational agents that help laypeople obtain health information or navigate services~\cite{esmaeilzadeh2025using,laymouna2024roles}, systems that support caregiver-provider communication and coordination around episodes of care~\cite{yang2024talk2care,sisk2026trust}, or AI and LLM-based decision-support tools with diagnostic reasoning~\cite{staes2024design,vrdoljak2025evaluating,wang2025preliminary}.} While these efforts provide important insights into technology use at specific touchpoints, they typically isolate the touchpoint from other stages of the overall healthcare trajectory, making it difficult to understand how patients’ experiences and practices evolve across time as they move between home, online spaces, and clinical care.
}

\yuling{
Responding to this limitation, a growing body of HCI and health informatics research has adopted a patient journey-based lens to understand and design for patients' end-to-end experiences. For example, \citet{jacobs2016cancer} developed a cancer journey framework that characterizes breast cancer patients' responsibilities, challenges, and technology use from screening and diagnosis through treatment and survivorship. \citet{suh2020parallel} conceptualized \textit{parallel journeys} for patients with comorbid conditions, mapping how cancer care and psychosocial care trajectories intersect and where technology could support care. Patient-journey mapping and design methods have been used to visualize emotions, values, and unmet needs along the care path~\cite{bui2023patient, meyerhoff2022meeting}
For instance, \citet{he2021patient} developed a patient-journey map of COVID-19 home isolation to identify pain points and digital touchpoints during isolation and follow-up. Working with patients, \citet{timinis2025co} co-created reassurance journey maps that show when and why post-operative cancer patients seek reassurance while enrolled in remote patient monitoring programs.
}

\yuling{
However, even these trajectory-oriented studies largely examine journeys in relation to particular diseases, services, or care programs, and these studies usually treat digital technologies as one element within the broader care ecosystem, rather than as continuously present, conversational collaborators.
To date, there is still limited work that holistically and cross-temporally examines how a single, flexible, yet powerful technology, such as LLM, threads through the entire healthcare-seeking journey and dynamically reshapes patients' experiences, practices, and relationships with the healthcare system.
Our study contributes to this line of research by introducing a longitudinal perspective into LLM-powered healthcare-seeking scenario. We systematically investigate how patients incorporate LLMs into their multi-stage healthcare trajectories in real-world settings, revealing the multiple roles LLMs play across different stages and the mechanisms through which they shape patient experiences and behaviors.
}

\subsection{\yuling{LLM-Powered Healthcare-Seeking: Promising and Limitation}}
\label{sub:related_work:2.2}

\yuling{
\minorreview{AI-powered technologies for patients have been advancing in the healthcare domain for decades, evolving from early general-purpose information retrieval~\cite{luo2008medsearch, bernstam2005instruments},
to machine learning and deep learning driven decision making systems \cite{wang2021brilliant, sun2025traditional}, specialized symptom checkers~\cite{semigran2015evaluation, meyer2020patient} and largely-scripted medical chatbots~\cite{rosruen2018chatbot, xu2021chatbot}.}
While these technologies have democratized access to multiple healthcare support, they are typically limited by rigid interaction patterns and specific tasks, i.e., performing predefined functions with constrained workflows, struggling to accommodate the dynamic, evolving needs that patients experience. 
}

\yuling{In contrast, the recent advances in LLMs bring a fundamentally different healthcare-seeking paradigm. 
With the strong reasoning capabilities, flexible natural-language interaction, and general-purpose adaptability, multiple LLM–based agentic systems are reshaping the landscape of healthcare-seeking practices~\cite{qiu2024llm}.
\minorreview{In the fields of HCI, AI, and digital health research, a growing number of LLM-powered health applications have been designed, developed, and deployed to support diverse clinical and non-clinical needs, ranging from supporting clinical practice like documentation or diagnosis~\cite{van2024adapted, liu2025generalist, goh2025gpt,yan2025multimodal,yang2025application}, accelerating medical research ~\cite{liu2025drbioright, li2025large}, to enhancing medical education~\cite{cascella_evaluating_2023, thirunavukarasu2023large}, and empowering patient follow-up and chronic disease management~\cite{li2025unveiling, kim2025optimizing}.}
These studies have significantly advanced the development of the LLM-powered healthcare sector.}

\yuling{Despite these promising developments,
to date, very limited studies have examined LLMs' role and impact through longitudinal usage in real-world clinical or everyday healthcare contexts.
Most existing studies on LLMs' role remain in an exploratory stage, largely evaluated on episodic snapshots of interactions in lab settings or hypothetical scenarios with interviews~\cite{song2025typing,foo2025benefits}, or retrospective analyses of online posts and focus group discussions~\cite{jung2025ve}.
Such examinations often fail to capture patients' dynamic needs, and also overlook how LLMs may vary in effectiveness~\cite{sallam2023chatgpt, nazi2024large}, safety~\cite{ahn2023exploring, sinha2023applicability, jeblick2024chatgpt}, reliability~\cite{rao2023evaluating, ahmad2023creating}, or ethical implications~\cite{templin2024addressing, zack2023coding} across different stages of the patient journey.
Integrating LLM technologies into healthcare, while simultaneously ensuring their usability, safety, and ethical deployment, remains a complex and challenging endeavor \cite{busch2025current,sallam2023chatgpt, nazi2024large}. 
The lack of longitudinal, in-situ evidence is particularly problematic, as it limits our ability to understand how these systems behave in real-world use, potentially leading to mismatches between design assumptions and actual patient needs or clinical realities \cite{sun2025traditional}. }

The concurrent rise of LLMs' powerful capabilities and potential risks creates a complex landscape for the adoption of LLMs in healthcare. Yet, while the academic and clinical communities proceed with caution, the general public has already begun leveraging these tools for health inquiries, blurring the boundaries of daily well-being and healthcare~\cite{xiao2024understanding,song2025typing, jung2025ve}.
In this work, we \yuling{contribute to this area by investigating the real-world dynamics of the interactions between patients and LLMs. 
Using a longitudinal approach, we specifically investigate the patterns, roles, and impacts of LLMs across the entire patient healthcare-seeking journey, and the methods patients develop to make sense of and manage their engagement with LLM technologies.}

\subsection{\yuling{Socio-Technical Dynamics in LLM-Powered Healthcare-Seeking}}
\label{sub:related_work:2.3}

In public narratives, whether in industry discourse, policy rhetoric, popular media, or academic publications, LLMs has often been envisioned as a powerful and nearly omnipotent technology (e.g., \cite{wikipediaLLM, govuk, raza2025industrial, oxfordinsights, eloundou2023gpts, bubeck2023sparks}) capable of undertaking highly challenging, complex, and value-laden tasks, ranging from creative content production \cite{haase2023artificial, ippolito2022creative}, to knowledge-intensive professional services \cite{bubeck2023sparks, eloundou2023gpts}, to decision support in high-stakes domains such as healthcare \cite{singhal2023large, shool2025systematic, xu2024mental}. In these accounts, LLMs are frequently celebrated as a key driver of human progress and innovation.

Yet, the introduction of an LLM-powered application into practical application settings is far from sufficient on its own. On the one hand, research has shown that despite the rhetoric of intelligence and automation, its practical deployment and effective use require extensive human labor \cite{Elish_2021_HiddenWorkAI}, such as configuring \cite{ulloa2022invisible}, maintaining \cite{chen2023maintainers}, contextualizing \cite{sun2023data}, and even reinterpreting AI outputs \cite{sun2025traditional}. Human involvement thus remains indispensable \cite{riek2025future}. On the other hand, as LLM becomes increasingly integrated into human society, the relationship between humans and LLM grows inherently complex. Recently, researchers have examined the social \cite{sun2025rethinking}, psychological \cite{sun2023care}, and ethical impacts \cite{chen2025moral, lei2025ai} of LLMs in domains such as education \cite{chen2025characterizing, sun2024exploring}, healthcare \cite{li2025large, mirzaei2024clinician}, and mental well-being \cite{sun2023care}. These studies converge on a shared conclusion that AI cannot, and should not, replace human agency. Instead, they call for the design and deployment of AI in ways that foster sustainability, safety, ethics, fairness, and broader social responsibility.

In the healthcare domain, this issue is particularly salient. While LLM-powered applications are widely promoted as powerful tools \cite{xu2024mental,yang2024drhouse,oh2024llm,clusmann2023future}, their effective adoption and use in practice typically depend on extensive human labor. For instance, prior studies have described how infrastructuring labor is required to integrate AI into inherently complex and fragmented healthcare practices \cite{gui2019making, verdezoto2021invisible, bhat2023we, wilcox2023infrastructuring}, how data work is needed when adopting AI into highly contextualized and human-centered healthcare practices \cite{sun2023data, sun2025traditional}, and how the everyday efforts of clinicians, technicians, and administrators are necessary to ensure the safety, reliability, and interpretability of LLM outputs \cite{ulloa2022invisible, procter2023holding, mirzaei2024clinician}. Additionally, studies have documented a range of psychosocial, safety, and ethical implications \cite{mirzaei2024clinician, haltaufderheide2024ethics} brought by LLM-powered healthcare applications, arguing for more human-centered technological directions \cite{ehsan2022human, deva2025kya}.

As a response, recent research has called for more empirical understandings of LLMs' role in practice, urging the HCI and related communities to move beyond technological promises to deeply examine the socio-technical dynamics of AI implementation, highlighting the need to situate AI within broader socio-technical, sociocultural, and socio-economic ecologies \cite{wang2021brilliant, sun2025traditional}. Joining this growing body of research, our study examines LLMs in real-world healthcare-seeking settings, foregrounding the complex interplay between AI capabilities and human practices.

\section{Methods}
\label{sec:methods}

To investigate the long-term impact of LLMs on patients' healthcare-seeking journeys, we conducted a four-week diary study to capture in-the-moment experiences and evolving practices, followed by in-depth, semi-structured exit interviews to elicit deeper retrospective reflections.

\subsection{\review{Study Context}}

\yuling{Our study was situated in China, a country that has long faced the challenges of insufficient and unevenly distributed medical resources~\cite{chen2021ten}.
As one of the world's most populous nations, China carries an enormous demand for healthcare services~\cite{yin2022hypertension}, while high-quality healthcare resources remain heavily concentrated in large public hospitals of major cities. 
This structure has made the issues of difficulty and high cost of seeing a doctor, a long-standing and significant public concern in Chinese society.
At the same time, China's family doctor system is still developing, and the primary care infrastructure has limited capacity to effectively divert patient flow. Most patients continue to rely on hospital-based registration and consultation, leading to overcrowded hospitals where patients commonly experience intense competition for appointments, long waiting times, and very short consultation durations \cite{liang2021patient, yan2021patient}. Healthcare providers, in turn, face excessive workloads and high rates of burnout \cite{zheng2022burnout, guo2021prevalence}.
Within such a strained healthcare system, patients often lack continuous channels for healthcare consultation, as well as adequate support for accessing reliable health information, making sense of diagnoses, or managing their conditions over the long term. Given such a context, LLMs, with their immediacy, low access barrier, scalability, and capacity for conversational engagement, position them as particularly valuable, holding substantial practical promise to alleviate healthcare burden and provide patients with more accessible healthcare support. 
Moreover, China is one of the most active countries globally in promoting the adoption of AI in healthcare. These factors create a highly relevant, dynamically evolving, and globally significant context for examining how LLMs can be integrated into patients’ real-world healthcare-seeking practices and what roles they can play.
}

\subsection{Participants and Recruitment}

Our recruitment criteria focused on participants who (1) were currently managing an active health concern, (2) anticipated making multiple in-person medical visits in the coming month, and (3) had prior experience using LLMs for broad health and well-being purposes. Our goal was to have a medically diverse participant pool to ensure the inclusion of a broad spectrum of patient experiences.
Recruitment started with an online form distributed via Chinese social media and public online communities, wherein we provided detailed information about the research team, study objectives, and eligibility requirements, as well as collected basic information from potential participants, including age, gender, health status, anticipated medical visits in the coming month, and AI Literacy. 
From the initial responses, we identified 75 participants who met the criteria and conducted screening calls with potentially eligible individuals. During these calls, we confirmed their health status, upcoming medical plans, and willingness to participate.

Our initial recruitment involved 42 participants. During the four-week study, seven participants withdrew or were lost to contact.
10 participants were excluded as their diary entries were consistently too brief or irrelevant to their health journey to provide meaningful data.
The final sample consisted of 25 participants (21 female, 4 male). Their ages ranged from 19 to 56 (mean = 25.2, SD = 6.9). 23 have bachelor's degrees, one has a master's degree, and one has a high school degree. \review{We also measured patients' AI literacy using the AI Literacy Scale~\cite{wang2023measuring}, with details presented in Table~\ref{tab:AI_Literacy}. Our user group demonstrated a moderate to high level of usage and familiarity with AI (4.8 to 6.6 out of 7, mean = 5.8, SD = 0.5).}
Their health conditions covered a broad spectrum of diseases, ranging from acute injuries (\eg bone fractures, infections) to the management of long-term chronic conditions (\eg diabetes, chronic pain) and serious diagnoses (\eg cancer). 
Table~\ref{tab:Demographic} lists their demographics and health condition details.

\begin{table*}[t]
\centering
\caption{The demographics, \review{proficiency in using AI in daily lives}, and health conditions of 25 participants. We also included the specific LLMs they used, together with the number of clinical visits during the four-week study period. Other than the well-known ChatGPT, participants used DeepSeek, Kimi, Qwen, and Doubao that are popular general-purpose LLMs in China. Others are health domain-specific LLMs developed specifically for health applications. \review{More details of these models can be found in Appendix Table~\ref{tab:LLM_Tools} and more details of participants' AI Literacy can be found in Appendix Table~\ref{tab:AI_Literacy}}.
}
\renewcommand{\arraystretch}{1.2}
\resizebox{\textwidth}{!}{%
\begin{tabular}{    
>{\centering\arraybackslash}p{0.6cm}   %
>{\centering\arraybackslash}p{0.6cm}
>{\centering\arraybackslash}p{0.6cm}
>{\centering\arraybackslash}p{2.2cm}
>{\raggedright\arraybackslash}p{5.5cm} 
>{\centering\arraybackslash}p{0.9cm}
>{\raggedright\arraybackslash}p{7.8cm} }
\toprule
\textbf{ID} & \textbf{Sex} & \textbf{Age} & \textbf{\review{Daily AI Use\textsuperscript{*}}} & \textbf{Health Conditions} & \textbf{Visits} & \textbf{LLM Used} \\
\midrule
P01 & F & 22 & \progressratio[7]{6}{1}{7} & Irregular Migraine & 0 & Kimi, Doubao, DeepSeek \\
P02 & F & 25 & \progressratio[7]{6}{1}{7} & Ectopic Pregnancy & 11 & Doubao \\
P03 & F & 21 & \progressratio[7]{7}{1}{7} & Post-cholecystectomy Syndrome & 3 & DeepSeek, Doubao \\
P04 & F & 21 & \progressratio[7]{6}{1}{7} & Chronic Menstrual Irregularities & 8 & Doubao, Kimi, AQ \\
P05 & F & 30 & \progressratio[7]{6}{1}{7} & Chronic Dyspepsia, Chronic Nausea & 3 & Kimi \\
P06 & F & 23 & \progressratio[7]{6}{1}{7} & Breast Tenderness, Ovarian Endometrioma & 1 & Doubao \\
P07 & F & 24 & \progressratio[7]{6}{1}{7} &  Chronic Gastritis, Otitis Media & 1 & ChatGPT, Doubao \\
P08 & F & 25 & \progressratio[7]{6}{1}{7} &  Kidney Stones & 11 & Doubao \\
P09 & F & 27 & \progressratio[7]{7}{1}{7} &  Thyroid Cancer & 2 & Kimi, DeepSeek \\
P10 & M & 22 & \progressratio[7]{5}{1}{7} &  Chronic Knee Pain & 1 & Doubao \\
P11 & M & 22 & \progressratio[7]{6}{1}{7} & Intermittent Dizziness, Arm Numbness & 3 & Doubao \\
P12 & F & 22 & \progressratio[7]{7}{1}{7} & Chronic Nausea & 0 & Doubao, iFlytek Xiaoyi, JD Health, AQ, ChatGPT, Zuoshouyisheng, Baidu Lingyi Zhihui, DeepSeek \\
P13 & F & 21 & \progressratio[7]{7}{1}{7} & Perforated Eardrum & 13 & Doubao, DeepSeek, Zuoshouyisheng, AQ, iFlytek Xiaoyi, JD Health, Xiaohe Health, Baidu Lingyi Zhihui \\
P14 & M & 22 & \progressratio[7]{7}{1}{7} & Fatty Liver, Elevated Transaminases & 2 & Doubao \\
P15 & F & 23 & \progressratio[7]{6}{1}{7} & Chronic Stomach Pain, Left Leg Fracture & 3 & Doubao, iFlytek Xiaoyi, Qwen, AQ, DeepSeek, Kimi, JD Health, Zuoshouyisheng \\
P16 & F & 21 & \progressratio[7]{6}{1}{7} & Chest Pain, Prone to Palpitations & 0 & Doubao \\
P17 & F & 19 & \progressratio[7]{6}{1}{7} & Sinusitis, Tooth Inflammation & 7 & Doubao, Xiaohe Health, AQ \\
P18 & F & 28 & \progressratio[7]{7}{1}{7} & Vaginitis & 8 & iFlytek Xiaoyi, DeepSeek \\
P19 & F & 29 & \progressratio[7]{7}{1}{7} & Endometrial Hyperplasia & 2 & Doubao \\
P20 & F & 24 & \progressratio[7]{5}{1}{7} & Knee Effusion and Chronic Pain & 4 & Doubao \\
P21 & F & 30 & \progressratio[7]{7}{1}{7} & Neurodermatitis, Seborrheic Dermatitis & 3 & iFlytek Xiaoyi \\
P22 & M & 26 & \progressratio[7]{6}{1}{7} & Diabetes & 3 & Doubao \\
P23 & F & 26 & \progressratio[7]{6}{1}{7} & Post-Hepatic Cirrhosis Infection & 10 & DeepSeek, JD Health, AQ \\
P24 & F & 56 & \progressratio[7]{5}{1}{7} & Breast Cancer & 2 & Doubao \\
P25 & F & 22 & \progressratio[7]{6}{1}{7} & Right Leg Fracture & 7 & DeepSeek, Doubao, AQ \\
\bottomrule
\end{tabular}%
}

\vspace{1ex} %
{\footnotesize
 \raggedright %
 \textsuperscript{*}\review{This item asks: ``\textit{I can skillfully use AI applications or products to help me with my daily work}''. Score: 1 (Strongly Disagree) to 7 (Strongly Agree).}
 \par
}

\label{tab:Demographic}
\end{table*}

\subsection{Study Procedure}

The study was conducted in two phases: a four-week diary study and in-depth, semi-structured exit interviews. The whole study procedure took place virtually to ensure flexibility and accessibility for participants. 

\subsubsection{Four-Week Diary Study.}
\label{sec:DiaryQuestionnaire}

We first conducted a four-week diary study, during which we observed how participants used LLMs for any queries related to their health conditions, and recorded their usage behaviors, feelings, and experiences in the diary.
Such a diary study allowed us to capture participants' daily in situ use practices, experiences, and reflections.
\review{Adopting from similar studies~\cite{hong2020using}, our study spanned four weeks to capture multiple patient-AI and potentially patient-provider interactions.}
We designed a structured diary questionnaire to elicit detailed data on participants' experiences with both LLMs and human clinicians. The questionnaire was designed based on our research questions and contained the following components:

\begin{enumerate}
\item \textbf{Interaction with LLM(s):} 
Participants reported their rationale for interacting with any LLMs, the content of their queries (via URLs or screenshots), and their assessment of the interaction. This included the perceived role of the tool, overall satisfaction, its impact on their decisions, and any feelings of trust, positive emotions, or frustration.
\item \textbf{Interaction with human clinician(s):} \review{If participants visited clinicians, participants entered additional diary data to report the reason for the visit (in-person or online), the topics discussed, the clinical outcomes (\eg prescriptions, procedures), and their overall experience.}
\item \textbf{Interaction with both:} If participants consulted both sources, they were asked to explain their reasoning, their perception of the relationship between the two, and how they navigated any different or conflicting information.
\end{enumerate}

All questions were designed to be general so that they could be inclusive and encouraged participants to share various levels of experiences and perceptions (see detailed diary questions in Appendix Table \ref{tab:DiaryQuestionnaire}).
To facilitate detailed responses, we provided an optional voice-to-text input for all open-ended questions.
On days with no relevant healthcare activities, participants were instructed to simply note ``No consultation today''.

The diary study began with an online introductory session three days prior to the main diary study for each participant, during which we introduced the overall study procedures.
Participants had the freedom to use any LLMs they were familiar with for any queries related to their health conditions.
We walked participants through the questions covered in our diary questionnaire to ensure they understood the meaning of each item. 

As part of the onboarding process, we provided participants with information about available LLMs and access methods, including general LLM tools as well as those focused on healthcare (see Appendix Table~\ref{tab:LLM_Tools}). We provided several examples to illustrate appropriate queries.
It is important to note that we emphasized in the session that these tools are supplementary and not a replacement for professional medical care.
Following this session, all participants signed an online informed consent form and received a text version of the user guide. More ethical considerations were detailed in Section~\ref{sub:methods:ethics}.

The diary study officially began after participants signed the consent form. During the four weeks, participants submitted daily diaries through the online questionnaire.
One author monitored the submission records. If more than three consecutive days of missing data were observed, we followed up with the participants to ensure compliance. Over the four-week diary study, 25 participants provided a total of 587 entries that contained consultation responses (Min = 8, Max = 28 per person). Within these responses, 478 involved only consultations with LLMs, 80 included both clinicians and LLMs, and 29 logs were solely with clinicians.
\yuling{It is noteworthy that, to capture real-world conditions, we maintained a strict no-intervention protocol. 
Throughout the whole diary study period, we provided no corrections or guidance, irrespective of the LLM-generated responses.}

\subsubsection{Exit Interview.}

Following the four-week diary period, we then conducted semi-structured interviews with each participant, aiming to further probe into their diary entries, clarify emerging themes, explore their experiences in greater depth, and understand the broader context of their healthcare journey. The interview protocol covered the following three main parts:
(1) We began by asking about their overall journey with the LLM tools, significant interactions they recalled, and any evolution in their perception of the LLMs' role over time.
(2) We then used specific diary entries as prompts, asking participants to elaborate on key decision points, ambiguous entries, or critical health incidents that required further clarification.
(3) Finally, we explored their perceptions of using LLM tools and human clinicians in tandem, focusing on how interacting with LLMs would affect their communication with human clinicians (and vice versa), and how that would affect their overall healthcare-seeking journey.
While following a semi-structured protocol, we intentionally kept the conversation open to probe emergent themes and elicit rich details from the participants' experiences.

All interviews were conducted via online meeting, lasting 30 to 90 minutes. With participants' permission, all interviews were audio-recorded and later transcribed for data analysis. 
After the interview, participants were compensated for their involvement in the whole study. They received \$1.5 for each valid diary entry (excluding those marked ``No consultation today'') and a final compensation of \$10 for completing both the interview and the entire study.

\subsection{Data Analysis}

We adopted an inductive thematic analysis approach \cite{fereday2006demonstrating} with a longitudinal lens to analyze the diary entries and follow-up interviews. Given the multi-dimensional nature of our datasets (25 participants' 587 diary entries, and in-depth interviews), our analysis focused not only on identifying recurring patterns but also on tracing how the roles of the LLMs dynamically evolved across participants' healthcare-seeking journeys.

Three authors participated in the initial phase of analysis. After familiarizing ourselves with the data, we independently conducted open coding on a subset (seven participants' diaries, 165 entries in total, and interviews), to surface initial codes.
Weekly meetings were held to discuss disagreements and refine the emerging codes.

Through these processes, we generated an initial codebook, primarily capturing the diverse roles LLMs played, participants' interaction patterns, and their experiential accounts of interacting with both LLMs and human clinicians. Guided by this codebook, two authors conducted the subsequent coding of the full dataset, meeting regularly to discuss discrepancies and iteratively refine the coding scheme. This process yielded a consolidated set of codes, which were then inductively clustered into higher-level categories, reflecting dimensions such as the roles of behavioral guidance, informational support, and emotional companionship that LLMs played, the interaction and mediation among patient-LLM-provider, as well as participants' evolving perceptions of trust, agency, and power dynamics.
Importantly, our analysis emphasized the temporal and processual nature of these roles and interactions. That is, rather than treating each diary entry as an isolated data point, we traced how the functions of LLMs shifted over time within each participant's healthcare-seeking journey. Through this trajectory-oriented perspective, we were able to situate the evolving roles of LLMs in relation to participants' changing health conditions, emerging needs, and shifting relationships with human clinicians, as well as across different stages of their experiential journey.

\review{Building on this preliminary code list, we re-focused our analysis at the broader level of themes, examining the contextualized meanings of emergent concepts and reviewing related literature to uncover their connections.}
Three authors iteratively consolidated our codes into an overarching theme, and carefully compared the identified themes to the dataset and refined the themes with the goal of ensuring internal homogeneity and external heterogeneity. 
After several rounds of discussion and refinement, we developed a thematic map consisting of four primary themes, each representing a distinct role played by LLMs and their corresponding dynamic processes in patients' healthcare-seeking journey. 
In the following section, we present these themes, using representative quotes and illustrative cases, which were translated from their original Mandarin into English by the authors. To protect our participants' identities, we use P01, P02, etc. to denote different participants.

\subsection{Ethical Considerations}
\label{sub:methods:ethics}
\minorreview{All study procedures were approved by our university's Institutional Review Board (IRB \#HR1-0367-2025).} We prioritized participants' well-being and data privacy throughout the research process. We obtained informed consent from all participants during the onboarding session after thoroughly explaining the study's purpose, procedures, potential risks, and their right to withdraw at any time without penalty.
~\review{All participants were informed via the consent form that all relevant data would be collected through the online questionnaire for scientific research purposes. Notably, the questionnaire platform we employed, Yibiaoda, strictly prohibits any unauthorized use or disclosure of collected data, thereby safeguarding participants' privacy~\cite{Yibiaoda}.}
A key ethical challenge was the potential for participants to misinterpret information from LLMs as professional medical advice.
To mitigate this risk, we explicitly and repeatedly emphasized during the onboarding session and in the written guide that these LLM tools are supplementary aids and not a substitute for consultation with qualified human clinicians.
All collected data, including diary entries and interview transcripts, were de-identified and stored on a secure server accessible only to the research team. In our analysis and reporting, we use PID and take care to remove any personally identifiable information from quotes or descriptions to ensure participant anonymity.

\section{Findings}
\label{sec:results}

Our findings suggest that in patients' long-term healthcare-seeking \yuling{trajectory, LLMs are far from being a single-function tool, e.g., information retrieval, decision making, or health Q\&A, as many current studies assume \cite{qiu2024llm, zhao2023survey}.} Instead, they continuously adapt their roles and dynamically integrate into patients' everyday health practices, shaping patients' healthcare-seeking experiences at the \textyellow{\textbf{behavioral}}, \textblue{\textbf{informational}}, \textred{\textbf{emotional}}, and \textgreen{\textbf{cognitive}} levels. Table \ref{tab:RoleDefinition} presents the definitions of the four roles adopted by LLMs, providing an overview of our key findings. \yuling{These four roles do not function as a sequential, time-series progression; rather, they form an interwoven dynamic, effectively reshaping the patient-doctor relationship in terms of agency, trust, and power.}

\begin{table*}[t]
\centering
\caption{The table lists the four roles of the LLMs in the patient's healthcare-seeking journey and their definitions, along with their specific manifestations as introduced in the respective sections of the findings.}
\label{tab:RoleDefinition}
\small
\renewcommand{\arraystretch}{1.3}
\resizebox{\textwidth}{!}{%
\begin{tabular}{ m{4.0cm} L{9.15cm} L{5.8cm} }
\toprule
\textbf{Roles of LLMs} & \textbf{Role Definition} & \textbf{Specific Manifestation} \\
\midrule
\multirow[c]{3}[8]{4.0cm}{\raggedright\arraybackslash \textyellow{\textbf{(Behavioral)}} Supporting and Negotiating Healthcare Decisions}
    & \multirow{3}[8]{9.15cm}{The Behavioral role involves the LLMs supporting patient decision-making across multiple healthcare-seeking stages, fostering a complementary partnership with clinicians, and enabling patients to actively calibrate trust and negotiate their healthcare choices.}      & LLM–Powered Decision Negotiation Across Stages (Section \ref{finding1.1}) \\
    \cmidrule(lr){3-3}
    &                                 & Complementary Partnership with Clinicians (Section \ref{finding1.2}) \\

    \cmidrule(lr){3-3}
    &            & Trust Calibration and Power Reconfiguration (Section \ref{finding1.3})  \\

\midrule
\multirow[c]{3}[8]{4.0cm}{\raggedright\arraybackslash \textblue{\textbf{(Informational)}} Facilitating Patient-Provider Communication}
    & \multirow{3}[14]{9.15cm}{The Informational role involves the LLMs facilitating patient–provider communication by enabling patients to come into the consultation ready to articulate their health information (e.g., conditions, symptoms), focus the clinical discussion effectively, and sustain their health agenda beyond the encounter.}      & Strengthening Self-Expression and Preparation Before Clinical Encounters (Section \ref{finding2.1}) \\
    \cmidrule(lr){3-3}
    &                                 & Guiding Communication Focus and Efficiency During Clinical Encounters (Section \ref{finding2.2}) \\

    \cmidrule(lr){3-3}
    &            & Empowering \& Sustaining Patient Agenda-Setting Across/After Encounters (Section \ref{finding2.3})  \\
\midrule
\multirow[c]{3}[5]{4.0cm}{\raggedright\arraybackslash \textred{\textbf{(Emotional)}} Providing Emotional Support and Companionship}
    & \multirow{3}[5]{9.15cm}{The Emotional role involves the LLMs acting as a dynamic companion that provides everyday reassurance while also serving as an emotional buffer during critical moments, with its overall impact on patient psychological well-being being highly situated and context-dependent.}      & Everyday Companionship and Care 
    
    (Section \ref{finding3.1}) \\
    \cmidrule(lr){3-3}
    &                                 &Buffering Anxiety in Critical Moments 
    
    (Section \ref{finding3.2}) \\
    \cmidrule(lr){3-3}

    &            & Situated Emotional Effects (Section \ref{finding3.3})  \\
\midrule
\multirow[c]{3}[0]{4.0cm}{\raggedright\arraybackslash \textgreen{\textbf{(Cognitive)}} Scaffolding Patients’ Sensemaking and Knowledge}
    & \multirow{3}[0]{9.15cm}{The Cognitive role involves the LLMs acting as an interpretive partner that scaffolds patients' knowledge by making implicit medical advice explicit and actionable, and by extending their sensemaking beyond the clinical encounter into daily life.}      & Making the Implicit Explicit and Understandable (Section \ref{finding4.1}) \\
    \cmidrule(lr){3-3}
    &                                 & Sensemaking Beyond the Clinical
    
(Section \ref{finding4.2}) \\

\bottomrule
\end{tabular}%
}
\end{table*}

\subsection{\textyellow{(Behavioral)} Supporting and Negotiating Healthcare Decisions}
\label{finding1-behavioral}

A central way in which our participants engaged with LLMs was through the ongoing scaffolding and negotiation of healthcare decisions. Our findings suggested, rather than providing single-point answers, LLMs accompanied patients across multiple stages of their trajectories, actively shaping how they acted on decisions, constructed the role boundaries between LLMs and clinicians, and gained greater agency.

\subsubsection{LLM–Powered Decision Negotiation Across Stages}
\label{finding1.1}

Our analysis shows that participants' healthcare decision-making processes were collaboratively enacted by LLMs, clinicians, and themselves. 
Empowered by LLMs, patients were no longer passive recipients, but actively engaged in the decision-making process at different stages of their healthcare-seeking journeys.

\paragraph{Before Clinical Encounter: Analyzing Health Conditions and Identifying Critical Healthcare Signals}
Before visiting a human doctor, participants often relied on LLMs to help them analyze their health conditions, identify signals warranting medical attention, and compare potential treatment pathways. For example, after experiencing inflammation in the ear, P13 consulted both Doubao and DeepSeek about the possible causes and risks. When both LLMs highlighted the severity of the problem and recommended an immediate ``otoscope examination'', she promptly followed their advice and sought medical care. In her 3rd diary entry, she wrote: ``\textit{The role of LLM is quite obvious. I asked it about my ear condition, and it immediately pointed out the severity, saying it might affect the cranial nerves. After hearing that, I immediately decided to go to the hospital for an otoscope exam.}'' Similar descriptions were also reported in the diaries of participants with other types of diseases, such as P03 (chronic conditions of post-cholecystectomy syndrome) and P06 (an acute case of vaginitis).%

\paragraph{During Encounter: Supporting Patient Agency in Treatment Decision}
During clinical encounters, the analyses and treatment suggestions offered by LLMs enabled patients to gain more knowledge about their conditions and exercise greater agency in discussing and shaping treatment plans with clinicians, rather than merely accepting their advice.
\review{This was particularly salient in the Chinese healthcare system, where outpatient encounters in large public hospitals are typically brief and highly compressed, and patients may see different clinicians across visits.}
For instance, P09, a participant with thyroid cancer, consulted the LLMs in detail about possible surgical options and their risks before undergoing surgery. On Day 18, she wrote: 
\begin{quote}
``\textit{The LLM carefully explained the pros and cons of three surgical options and compared... This gave me a clearer understanding of each option's advantages and disadvantages, and provided important reference points for choosing my surgical plan.}''
\end{quote}

In some cases when clinicians were unable to determine the cause or treatment, patients engaged in the attempt of self-diagnosis with the help of LLMs. For example, P18 visited a doctor for fever, but blood tests did not reveal the cause.
She noted: ``\textit{AI...listed a large number of possible causes. I began ruling them out one by one by myself.}''
Although P18 could not obtain a definitive diagnosis of the fever even after recovery, LLMs provided support for her to explore multiple possible causes.

\paragraph{After Encounter: Guiding Follow-Ups and Health Monitoring}
After the medical treatment, patients also relied on LLMs to, for instance, schedule follow-ups (\eg  P02, P08, P17, P18, P23), decide whether a revisit was necessary (\eg P08, P17, P23, P25), prevent recurrence (\eg P02, P08, P10, P18), etc. For instance, according to P02's diaries, after an ectopic pregnancy surgery, P02 consulted LLMs every day about her physical recovery, as well as lifestyle-related concerns such as diet, sleep, and exercise. The LLM guided her postoperative rehabilitation and help distinguish normal symptoms from those requiring immediate care. On Day 10, she wrote: 
\begin{quote}
    ``\textit{The LLM's responses were well-organized and clear. It differentiated normal recovery patterns from abnormal conditions, helping me identify when I needed to seek medical care promptly. This prevented delays... and improved the efficiency.}''
\end{quote}
At the later stage of the study, P02 visited her doctor on Day 17 and 18 to have postoperative medical check-ups.

\subsubsection{Complementary Partnership with Clinicians}
\label{finding1.2}

During the continued use of the LLMs, participants gradually developed a self-understanding of the role relationship between LLMs and clinicians.

\paragraph{Role Complementarity}
Many participants pointed out that LLMs' responses were usually ``broad and comprehensive'' and could quickly provide multiple possible causes and general medical knowledge, but lacking an understanding of the specific individual and their full diagnostic and social context. By contrast, clinicians were perceived as more capable of integrating complete diagnostic information with social–environmental factors, thereby offering direct, personalized, and precise diagnoses and prescriptions.
\review{However, participants acknowledged that this complementarity was influenced by the structural reality of the local healthcare system, necessitated by the resource constraints and limited attention from clinicians in public hospitals.}
As P02 noted in her 6th entry:

\begin{quote}
    ``\textit{It [the LLM] mentioned many possible causes of dizziness, such as anemia,  which might be accompanied by paleness and palpitations. I learned, but couldn't determine whether I fit the typical case. The doctor directly said my dizziness was caused by post-surgical weakness and orthostatic hypotension, and advised me to sit for 3 minutes before standing up next time.}''
\end{quote}

\paragraph{Boundary Construction}
Within this complementary relationship, participants also gradually constructed functional boundaries between LLMs and clinicians in their treatment journey, \review{which was often drawn based on the availability of medical resources.} Participants trusted clinicians to set the precise professional direction, such as diagnostic conclusions and treatment plans, while LLMs translated these into more easily accessible, concrete, everyday behavioral suggestions, \review{a role ideally filled by nursing staff or primary care providers, which are often under-resourced in the local healthcare setting.} For instance, P08, a kidney stone patient, explained in the exit interview:
\begin{quote}
    ``\textit{Doctors won't constantly track your condition, and they don't know every little detail of your daily life. They mainly give you a general direction. But LLM can. In the later stages, I treated LLMs as my health assistant, which could turn the human doctor's advice into specific daily actions I needed to do.}''
\end{quote}

Beyond translating medical conclusions into daily practical actions, participants suggested that LLMs often played a primary role in addressing ``small matters'' or ``minor illnesses'', sometimes even surpassing human clinicians. These ``small'' issues mentioned by our participants included precise medication use (\eg how many drops of eye drops from P13 and P15), choosing hospitals or departments (14 out of 25 participants), as well as advice on diet (24 participants), lifestyle (24 participants), exercise (10 participants), living environment adjustments (P06, P16, and P18), etc. For example, P24, a breast cancer patient undergoing chemotherapy, noted on Day 11:
\begin{quote}
    ``\textit{I felt better today... I asked the LLM many related questions, and it recommended TV shows and music that really suited my taste. From now on, I'll focus on eating and drinking well, keeping a good mood, and preparing for the next round of chemotherapy.}''
\end{quote}
These conversations about daily well-being advice were closely relevant to their healthcare conditions and treatment, showing that the roles of LLMs can extend into the patient's daily life to support a more holistic model of self-care.

\subsubsection{Trust Calibration and Power Reconfiguration}
\label{finding1.3}

In traditional patient-provider relationships, clinicians typically hold the dominant role, with patients passively accepting their advice.
\review{This asymmetric pattern has been particularly prominent in our study's context, where overcrowded hospitals, strong professional hierarchies, and limited time can reinforce such dynamics.}
However, we found that under LLM-powered healthcare-seeking, patients reconfigured these power dynamics: patients gained knowledge-based agency and equal conversational status to actively participate in decision-making jointly with clinicians.

For example, after receiving a diagnostic conclusion from a human doctor, patients could consult LLMs again. If the answers converged, they felt greater confidence in the doctor's advice and adopted that directly; if they diverged, patients entered a state of ``boundary tension'', comparing, weighing, and sometimes even challenging human doctor's professional opinions. 
For instance, P18, a patient with vaginitis, described her experience after a physician recommended red-light and ozone therapy:
\begin{quote}
    ``\textit{The LLM said it was useful, but suggested that one round (seven days) was enough; doing more could damage the mucosa. So I decided to do one round but no more within the month. If the doctor recommends more, I'll ask for alternative medication instead.}''
\end{quote}
P18 discussed this decision with her doctor in a follow-up conversation and confirmed that this was a viable option.

Some patients consulted multiple LLMs for cross-validation before making final decisions.
For instance, P23 noted on Day 7 in her diary: ``\textit{I'm a bit skeptical about relying on just one LLM, mainly because I'm not sure if its analysis is professional enough.}''
Patients could obtain more diagnostic directions and agendas from LLMs to discuss with their clinicians (see more details in Section \ref{finding2.3}). Through these practices, patients recalibrated their trust in clinicians and LLMs, and redefined their own agency and power within the healthcare process. 

In some cases, participants even noted that LLM-derived knowledge empowers them to make decisions when confronting urgent recommendations from clinicians. 
For instance, P19, a patient with uterine disease, mentioned in the exit interview:

\begin{quote}
``\textit{I feel that the initiative is in my own hands. The doctor said the surgery needed to be scheduled that day, otherwise there would be no available bed... But I thought I needed to think more about it, discuss it with the LLM, and discuss it with my family.}''
\end{quote}

Taken together, the results show that in longitudinal care, patients adopted LLMs to scaffold and negotiate decisions, jointly supporting decision-making with clinicians across different stages, thereby positioning themselves as more active agents and reshaping traditional power relationships in healthcare.
\review{We also highlight that such empowerment also comes with potential risks, as elaborated in Section \ref{sub:discussion:risks}.}

\subsection{\textblue{(Informational)} Facilitating Patient-Provider Communication}
\label{finding2-informational}

Patients' diaries and interviews revealed that LLMs also acted as scaffolds for patient-provider communication. This scaffolding strengthened patients' self-expression prior to consulting clinicians, guided the focus during the discussion, and empowered patients to sustain and expand their agenda beyond the clinical visits.

\subsubsection{Strengthening Self-Expression and Preparation Before Clinical Encounters}
\label{finding2.1}
In traditional clinical visits, patients often struggle to describe their conditions accurately to doctors, leading to vague or incomplete information. Our results show that the role of LLMs in this respect was significant. Specifically, 18 out of 25 participants pointed out that, LLMs guided them to produce accurate, clear, and professional descriptions of their symptoms (\eg ``\textit{LLM told me to describe changes in wound exudate and sensations during movement to the doctor.}'' (P02, Day 3)), reminded them to record key health indicators (\eg ``\textit{AI asked me to observe the color and texture of secretions with a cotton swab every day.}'' (P18, Day 13)), and prompted them to prepare relevant materials before the clinical encounter (\eg ``\textit{AI suggested I bring previous case records, test results, and allergy history.}'' (P20, Day 4)). 

Additionally, some participants noted that LLMs helped them rehearse diagnostic reasoning by highlighting key symptoms to recall and present.
The case provided by P13 in the interview could illustrate this role very well: 
\begin{quote}
    ``\textit{AI told me the typical symptoms of otitis media, such as pus discharge and buzzing or muffled sounds in the ear. I started to self-diagnose and thought it might be otitis media. When I asked the doctor, he said, `Are you sure you heard the obvious popping sound?' I said yes. Then the doctor immediately checked again and confirmed [the diagnosis].}''
\end{quote}

\subsubsection{Guiding Communication Focus and Efficiency During Clinical Encounters}
\label{finding2.2}

\review{In many countries, such as China where our study was conducted, healthcare resources are limited and patients' clinical encounter times are notoriously brief as emphasized by our participants (often lasting only a few minutes).}
In this context, in addition to preparing before the counter,
our participants further emphasized LLMs' significant roles during the encounter, \review{helping them structure their questions and maximize the value of the limited time with clinicians.}
As P05, a patient with gastric issues, wrote on Day 8:
\begin{quote}
    ``\textit{The LLM taught me how to precisely explain my condition to the doctor, including specific symptoms during flare-ups, dietary triggers, and physical states. [...This] improved efficiency.}''
\end{quote}

\review{In the highly efficient but hurried environment of outpatient clinics,} LLM could help patients convey key information efficiently. In other cases, LLMs also encourage our participants to enter clinics with targeted agendas rather than asking less relevant questions. For example, P17, a patient with periodontitis, wrote on Day 10 that she prepared three specific questions (``\textit{crown materials, postoperative chewing, and discoloration}'') based on the LLM's suggestions.

Additionally, patients also noted that LLMs guided them to be more proactive in seeking clarification. As P13 reflected in her Day 12 diary: ``\textit{Sometimes doctors won't give a direct answer. The final decision is still up to me... The LLM then told me how to guide the doctor into giving a more definite response.}''

\subsubsection{Empowering and Sustaining Patient Agenda-Setting Across/After Encounters}
\label{finding2.3}
Through continuous LLM scaffolding and reshaping their agency, many participants developed habits of entering clinics with clear agendas and sustaining active dialogue during follow-ups and cross-stage communication. As P06 wrote in her 14th diary about LLM-enabled self-monitoring and adjustment:
\begin{quote}
    ``\textit{LLM told me some very important points: physiological changes may last only 1–2 days, while pathological ones last more than three days and come with abdominal pain... So I judged it to be physiological.}''
\end{quote}

Other participants also described how LLMs inspired them to expand their medical agendas. For instance, P06, a patient with ovarian endometrioma, explained in the exit interview:
\begin{quote}
    ``\textit{I usually only had abdominal ultrasound and Ca19-9 for follow-ups... I hadn't thought about sex hormone panels. The LLM suggested this, and I realized it was more suitable and necessary for me to pay attention to.}''
\end{quote}

Taken together, LLMs facilitated patient-provider communication not only by preparing patients to articulate and organize their concerns before encounters, but also by helping them focus their questions during visits and sustain agenda-setting across stages.
Through these roles, patients no longer relied exclusively on human doctor-led Q\&A formats. Instead, they became capable of self-observation, preparation, and structured communication, which enhanced both efficiency and quality.
\review{These effects were amplified by the specific conditions of the Chinese healthcare system, where structural constraints on time and continuity make such communication support particularly salient.}

\subsection{\textred{(Emotional)} Providing Emotional Support and Companionship}
\label{finding3-emotional}

Emotional support and companionship are also central roles of LLMs. Our diary analysis shows that emotional support was one of the most frequently mentioned functions, with all participants acknowledging it and more than half of the daily entries referring to it. Nearly every participant recorded abundant descriptions of LLMs' emotional support, such as ``\textit{I felt reassured after asking}'', ``it reduced my anxiety'', and so on. 
Further, such an emotional supporting role is dynamic, situated in patients' healthcare-seeking journeys, \review{and deeply entangled with the local sociocultural and healthcare context, where stigma around certain conditions, perceived power distance with clinicians, and limited psychosocial support in clinical encounters can leave patients feeling isolated.}

\subsubsection{Everyday Companionship and Care}
\label{finding3.1}
Across long healthcare-seeking journeys, emotional fluctuation and anxiety are common for patients. Our study found that, with its broad knowledge and constant availability, the LLMs gradually became an everyday companion. Many participants noted that LLMs addressed their anxiety through providing clear, structured explanations of their health issues, which in turn provided a strong sense of assurance for them. As P06 mentioned in her 19th diary: 
\begin{quote}
    ``\textit{I always feel dizzy, and often get anxious because of the dizziness. Today, the LLM suggested that it might be due to low blood pressure and provided a complete breakdown from `why low blood pressure occurs' to `how to alleviate it', `possible impacts', and `when to seek care'... making me feel respected and guided with a sense of reassurance.}''
\end{quote}

\review{In contrast to the crowded, noise-filled, and efficiency-driven atmosphere of large public hospitals, LLMs offered a private, patient-centric space for emotional release. Participants frequently contrasted the efficiency required of clinicians against} LLMs' accessibility (always available for any questions) and conversation style (a certain degree of empathy).
As P20, a patient with knee disease, mentioned during the interview:
\begin{quote}
    ``\textit{I feel that the relationship between me and the LLM is quite equal. You can be extremely detailed and completely unreserved without worrying that your question might be dismissed...
    In contrast, some doctors in hospitals may show impatience.}''
\end{quote}

The interactive experience led participants to develop distinct role perceptions of the LLMs, aligning with their complementary partnership role with clinicians on the behavioral level (Section~\ref{finding1.2}).
For instance, they came to regard the LLM as ``\textit{a considerate companion, instead of a rigid robot that only recites preset rules or scripts}'' (P08), ``\textit{an old friend who understands my troubles, instead of a condescending doctor}'' (P05), and ``\textit{a friend who stays with me anytime, instead of just a tool for looking up information }'' (P13). Such experiences allow patients to express their confusion and pour out their anxieties anytime, anywhere, without hesitation. For instance, P08, who has only used LLM tools a few times before joining the study, noted in her 25th diary:
\begin{quote}
    ``\textit{At the beginning, I thought it [the LLM] was just a machine that only recites rigid rules. But my feelings are completely different [now]. Chatting with it feels like chatting with an old friend who understands my troubles, rather than listening to rigid preaching. It doesn't give me advice in a condescending way like many doctors do; instead, it considers my feelings from my perspective...
    I feel very warm and secure.}''
\end{quote}

\subsubsection{Buffering Anxiety in Critical Moments}
\label{finding3.2}

Beyond daily companionship, our study suggested that LLMs also provided patients with a significant ``emotional buffer'' during many critical healthcare moments, such as before clinical encounters, before surgeries, while awaiting test results, or when doctors are making major decisions, helping patients cope with uncertainty and intense stress. 
For instance, P04, who was dealing with gynecological issues yet stigma concerns,
wrote in her 14th diary:
\begin{quote}
    ``\textit{As a young person, I feared that doctors or others would hold stereotypical views of me, thinking I ended up needing gynecological checks because of reckless behavior. After talking to Doubao [an LLM], I felt encouraged... and went to get checked. }''
\end{quote}

Participants also noted that compared to the tension that may arise from direct cues like a doctor's facial expressions or tone of voice, LLMs were usually gentle and caring, which effectively helped them relieve intense psychological stress. As P06 mentioned during the interview:

\begin{quote}
    ``\textit{Sometimes when I visit a doctor in person, if the doctor frowns, my heart skips a beat. I get really nervous and feel awful. But the LLM is more gentle; it gives me psychological preparation and comfort... Whether the outcome is good or bad, I feel much more at ease.}''
\end{quote}

Additionally, many participants pointed out that LLMs offered patients crucial ``knowledge buffering'', which indirectly serves as an emotional buffer. They could help patients know what to expect, thereby alleviating their tension and anxiety. Our participants emphasized that this type of companionship was more effective in reducing stress. For example, P17 wrote on the day of her surgery (Day 8):

\begin{quote}
    ``\textit{I need to have surgery today, with general anesthesia, and I felt a bit flustered. I talked to the LLM and asked many questions. It patiently explained various things to me, including what preparations to make, potential risks, and post-surgery precautions. After our chat, I felt much more at ease. When I'm emotionally upset, my friends might comfort me with kind words, but the LLM offers not only verbal comfort, but also very specific advice.}''
\end{quote}

\subsubsection{Situated Emotional Effects (sometimes negative)}
\label{finding3.3}

While the aforementioned positive emotional impacts, our diary analysis also identified some negative emotional impacts brought by LLMs, which were primarily attributed to the functional limitations of LLMs.
For instance, the failure to accurately recognize user intent (\eg ``\textit{I wanted advice for at-home care, but it kept telling me to see a doctor right away.}'' (P13, Day 10)), short memory and inconsistent advice (\eg ``\textit{It can't remember what I asked yesterday, and its advice today often contradicts what it gave yesterday.}'' (P20, Day 3)), inability to provide clear answers (\eg ``\textit{It elaborates on with all sorts of analyses but never gives a clear conclusion.}'' (P23, Day 2)), etc. 

Some participants also felt that the LLM's formulaic responses lacked sincerity, which in turn reduced their trust in it. For instance, P13 mentioned in the interview: ``\textit{I often feel like it's a player\footnote{Player (\textit{haiwang}, 海王 \xspace in Chinese) refers to the person who maintains casual romantic relationships with multiple people at the same time.} who just repeats the same lines to everyone, then moves on to the next one right away.}'' Additionally, a small number of participants noted that the broad, all-encompassing knowledge presented by the LLMs might, to some extent, increase their anxiety. As P13 recorded in her 1st diary: ``\textit{It told me that an ear infection could affect the cranial nerves, and that drinking too much coffee would make it worse. It made my condition sound so serious.}''

Further, we also found that users' emotional perceptions of the LLM are strongly situated with the type of questions they asked. Relatively, patients gave more positive feedback when consulting about lifestyle-related issues such as diet, exercise, and rehabilitation, and more negative feedback when asking about more specialized, in-depth questions, for which clinicians may be more suitable.
Moreover, a patient's health status can also influence their emotional experiences toward LLMs. For example, P5 wrote in her 23rd diary: ``\textit{I wasn't feeling well today. I was already a bit impatient myself, and the LLM's answers were rambling and unsatisfactory in many ways. After asking, I felt even worse.}''

Taken together, our findings suggested that LLMs play a crucial role as emotional supporters throughout patients' healthcare journeys. It establishes a continuous, situated companionship with patients, enabling them to experience companionship, respect, care, or feel more anxious and frustrated.
Meanwhile, the potential negative experiences further suggest the importance of them being more contextually aware, as will be discussed in Section~\ref{sec:discussion}.

\subsection{\textgreen{(Cognitive)} Scaffolding Patients’ Sensemaking and Knowledge}
\label{finding4-cognitive}

During the practical healthcare-seeking process, patients often encountered difficulties in understanding professional terminology, missed important health or medical details, or felt hesitant to ask certain questions.
In these moments, the LLM acted as an interpretive partner, helping patients translate, extend, and contextualize the professional information they received from clinicians, and scaffolding their sensemaking and knowledge.

\subsubsection{Making the Implicit Explicit and Understandable}
\label{finding4.1}

\review{Because of the extreme time pressure in local clinical settings}, clinicians often provided concise, high-level guidance rather than detailed explanations. 
For instance, a doctor might mention that surgery was necessary without elaborating on the exact process (P17, Day 7), or prescribe medication without specifying usage details and precautions (P13, Day 4). 
\review{This ``high-efficiency'' communication style assumes a level of patient health literacy or compliance that often does not exist.}
In these situations, the LLM played a crucial role in translating difficult, abstract medical terminology into content that was both easy to understand and easy to act upon, while also filling in the details that clinicians left unspoken or that patients missed at the time.
For example, P17 used the LLM to learn the detailed steps of a laryngoscopy;
P13 and P03 asked about the exact dosage and method of using prescribed drugs; and P19 and P02 used it to clarify the implicit meanings in diagnostic reports. Participants emphasized that such detailed clarifications were essential for extending clinical advice into daily routines.

In these scenarios, the value of the LLM lies not only in supplementing the health knowledge but also in restructuring the format and narrative of knowledge itself to help patients better understand these concepts.
In doing so, the LLM made implicit professional assumptions explicit, enabling patients to move from structured understanding to actionable knowledge. \review{This decompression role was particularly important in the local healthcare system, where structural pressures often leave clinicians little time for detailed explanation and where patients may have limited access to health education resources outside the hospital.}

\subsubsection{Sensemaking Beyond the Clinical}
\label{finding4.2}

Diary data from our participants further revealed that beyond reconstructing complex medical knowledge, the LLM also served as a gap-filler in patients' healthcare-seeking journeys, providing nuanced, sensitive, and multi-dimensional details outside the scope of medicine, answering questions that clinicians might dismiss as trivial, patients might feel embarrassed to ask, or that were forgotten in the rush of clinical encounters.
For example, P8, after surgery, asked the LLM whether she could do housework; P05 raised sensitive gynecological questions with the LLM, free from the discomfort of being judged by doctors; and P09 even asked fortune-telling-like questions during moments of anxiety, seeking psychological comfort.
This scaffolding role meant that patients' sensemaking was no longer confined to clinical diagnosis and medical treatment alone, but extended into postoperative recovery, daily self-care, emotional adjustment, and lifestyle decisions.
This not only builds synergies with complementary roles with clinicians (Section~\ref{finding1.2}), but also provides additional evidence that the LLMs are blurring the boundary between daily well-being scenarios and healthcare.

Taken together, this section highlights the role of LLMs as powerful cognitive scaffolds in patients' healthcare-seeking journeys. 
By transforming implicit professional language into actionable guidance and filling in knowledge gaps beyond the clinic, LLMs enabled patients to apply medical advice to their daily lives, fostering holistic sensemaking that linked expertise with lived experience.
\review{These functions are deeply shaped by, and responsive to, the specific structural features of the local healthcare system (where primary care and community-based support infrastructures are still developing), while also pointing to forms of support that may be relevant in other resource-constrained or hospital-centric settings.}

\section{Discussion}
\label{sec:discussion}

Our study demonstrated that the role of LLMs goes far beyond providing medical information or supporting diagnoses; instead, they are deeply embedded in patients' healthcare-seeking journeys, continuously shaping decision-making, communication, emotional support, and sensemaking.
These findings reveal the diverse and situated roles that LLMs play in practice, which are often overlooked in existing discussions centered primarily on efficiency and accuracy.
\review{In what follows, we draw on our findings to examine more deeply how LLMs should be repositioned in healthcare and how they reconfigure patient–provider-AI relationships. We also discuss the important ethical risks associated with patients' adoption behavior of LLMs, ending by outlining design implications for future research.}

\subsection{Repositioning the Role of the LLM: \yuling{Multi-layered, Dynamic, and Interwoven Effects}}
\label{sub:discussion:boundary}

As mentioned in Section~\ref{sub:related_work:2.3}, much of the existing literature and public discourse positions LLMs primarily as powerful tools for specific health tasks such as information seeking, triage, and decision support, often with a focus on efficiency and accuracy in clinical workflows~\cite{xu2024mental, yang2024drhouse, xiao2024chinese, kim2024health}. \review{At the same time, emerging HCI and health informatics work has begun to foreground patients' perspectives on AI, including how they want decision power to be shared between clinicians and AI systems~\cite{kim2024much}, and how people appropriate general-purpose LLM chatbots for emotional and therapeutic support in mental health contexts~\cite{song2025typing, jung2025ve}. Together, these studies argue that LLMs are not merely computational tools, but socio-technical actors whose roles are negotiated, situated, and value-laden.}

\review{In particular, recent work on LLMs for mental health has shown how users craft idiosyncratic support roles for chatbots, filling gaps in everyday care and seeking validation, emotional containment, and preparatory support for therapy~\cite{song2025typing, jung2025ve}. Conceptual work in health informatics and AI ethics similarly points to LLMs as reshaping patient agency, access, and vulnerability in global healthcare~\cite{armoundas2025patient}. While these studies already move beyond a narrow ``information provision'' framing~\cite{chen2021ten}, they largely focus on mental health contexts or speculative decision scenarios, and typically examine single episodes or relatively short-lived interactions.}

\review{Our study extends and deepens these perspectives by showing how LLMs are woven into \emph{longitudinal, somatic} healthcare-seeking journeys that span multiple phases of care. Rather than focusing on mental health conversations in isolation or hypothetical decision vignettes, we trace how patients appropriate LLMs over time across symptom appraisal, clinical encounters, post-consultation sensemaking, recurrence monitoring, and everyday life. In this process, LLMs do not function only as diagnostic or decision-support tools; they emerge as continuous socio-technical partners whose roles are \emph{multi-layered} (behavioral, informational, emotional, cognitive), \emph{dynamic} (shifting as the trajectory unfolds), and \emph{interwoven} (each layer shaping and being shaped by the others). Our findings thus empirically substantiate, and move beyond prior claims that LLMs ``go beyond information provision'' by showing how they reorganize the temporal and relational aspects of healthcare-seeking itself.}

\review{
Building on these findings and prior scholarship, we argue for repositioning LLMs in healthcare: from task-bounded tools toward \emph{situated, dynamic boundary companions} embedded in patients' longitudinal journeys. This notion connects to, but is distinct from, earlier HCI concepts such as boundary objects and boundary negotiating artifacts in patient-generated data~\cite{chung2016boundary,leigh2010not}, and relational agents designed to maintain long-term therapeutic alliances~\cite{bickmore2005s, bickmore2010empathic}. Boundary objects and boundary negotiating artifacts emphasize how shared representations (e.g., self-tracking data) mediate collaboration between patients and clinicians~\cite{chung2016boundary}, while relational agents are carefully designed conversational systems that build rapport and continuity around specific health behaviors~\cite{bickmore2010empathic}. By contrast, the boundary companion we observe is \emph{not} a purpose-built artifact or domain-specific agent, but a general-purpose, open-ended LLM that patients themselves enlist, configure, and re-interpret across the trajectory, allowing it to simultaneously mediate data, decisions, relationships, and emotions.}
We unpack three key boundaries it traverses.

First, the LLM traverses \textbf{the interpersonal and informational boundary between patients and clinicians}. This boundary is traditionally marked by limited medical resources, power dynamics, and knowledge gaps. Our findings show LLMs acting as translators and communication facilitators (Section \ref{finding2-informational}, \ref{finding4-cognitive}). They helped patients deconstruct complex medical terminology, prepare structured questions, and rehearse symptom descriptions before appointments (Section \ref{finding2.1}). This process transformed patients from passive recipients into active, prepared participants, enabling them to challenge or collaboratively refine treatment plans with their clinicians (Section \ref{finding1.3}) and thus reconfiguring the traditional power dynamic between patients and doctors.

Second, it crosses \textbf{the contextual boundary between the clinical setting and the patient's everyday well-being}. While clinicians provide professional, high-level directives, LLMs translated these into the ``specifics of everyday life,'' such as how to prepare certain foods or break exercise into manageable steps (Section \ref{finding4.1}).
They addressed granular concerns—from postoperative care and diet to managing mood (Section \ref{finding1.2})—that fall outside the scope of a typical clinical visit. This role extends healthcare beyond the episodic encounter, embedding it into the continuous elements of a patient's life and supporting a more holistic model of well-being.

Finally, the LLM blurs \textbf{the functional boundary between cognitive decision-making and affective emotional support}. Our findings reveal that these two aspects are not separate but deeply intertwined domains. The LLM provided ``knowledge buffering'' (\eg explaining a surgical procedure in detail), which directly served as an ``emotional buffer'' by demystifying the unknown and reducing anxiety (Section \ref{finding3.2}). Receiving a clear, structured explanation (Section~\ref{finding4.1}) for a worrying symptom provided a sense of reassurance and control (an affective outcome) (Section \ref{finding3.1}). This fusion of roles demonstrates that effective health support is not just about delivering correct information and knowledge, but about delivering it in a way that fosters emotional well-being.

By fluidly navigating these interpersonal, contextual, and functional boundaries, the LLM emerges not as a static tool but as a dynamic companion that integrates disparate facets of the healthcare experience. Drawing on this conceptualization, we suggest that future research and design should recognize the potential of LLMs as boundary companions, shifting the focus from diagnostic accuracy or efficiency alone toward their longitudinal, situated, and multidimensional engagement in patients' health journeys.

\subsection{Reconfigured Patients' Agency, Trust, and Power}
\label{sub:discussion:reconfigure}

Our findings also reveal \yuling{LLMs are reconfiguring} traditional dynamics of patients' agency, trust, and power in healthcare, wherein clinicians often hold epistemic authority and patients are expected to adopt relatively passive roles, accepting diagnoses and treatment plans \cite{parsons2013social, kaba2007evolution}. The integration of technologies has undoubtedly reshaped this relationship. Previous HCI work, for instance, has documented how patients seek to gain more active roles through online health communities \cite{ma2018professional}, peer support \cite{hu2019peer}, or personal health tracking \cite{kaziunas2017caring}.

Our findings join this research thread, showing that LLMs enabled patients to claim greater agency and power in their healthcare journeys. For instance, our results showed that patients use LLMs in preparing themselves before clinical encounters (Section~\ref{finding2.1}), sustaining agendas across stages (Section~\ref{finding2.3}), comparing alternative treatment options, and even challenging clinicians' recommendations (Section~\ref{finding1.3}). This process reflects a redistribution of trust, as patients calibrated their confidence by cross-validating between LLMs' advice and clinicians' guidance, sometimes placing equal weight on both. In turn, this recalibration reconfigured power relations: patients, empowered with LLMs' scaffolding, no longer positioned themselves as passive recipients of expertise but as active negotiators capable of shaping healthcare decisions.

These reshaped relationships resonate with theoretical discussions of ``epistemic authority'' and ``knowledge asymmetry'' in medical sociology \cite{heritage2006communication}, but also extend them by showing how LLMs can act as socio-technical mediators that equip patients to contest, reinterpret, and negotiate medical authority. In this sense, the role of LLMs goes beyond providing knowledge to actively restructure the medical decision-making process, shifting the patient-provider relationship toward more dialogical and bilateral forms.

This newfound agency, however, is a double-edged sword that poses strong accuracy expectations and ethical risks for using LLMs in healthcare-seeking. Although patients can be more empowered with the support provided by LLMs, their lack of medical knowledge may hinder them from carefully distinguishing the errors and hallucinations introduced by LLMs, a problem of AI over-reliance identified by the HCI community~\cite{vasconcelos2023explanations,he2023knowing}. The danger lies not just in factual inaccuracies, but in the persuasive and seemingly empathetic veneer of authority that LLMs project. A well-organized, comprehensive, and reassuring answer (Section \ref{finding3.1}) might be trusted for its tone and structure rather than its clinical validity, leading to misplaced confidence. This creates a state of \textit{fragile agency}, where patients are empowered to act (\eg by challenging a doctor's advice), but this empowerment is built on a potentially unstable foundation. In such scenarios, the patient unknowingly shoulders a greater burden of clinical responsibility, making high-stakes decisions without the expertise to fully assess the risks. The LLM lowers the barrier to accessing medical information, but not to obtaining medical expertise, creating a new and complex landscape of patient risk.

We thus argue that future work on LLMs in healthcare must explicitly recognize and navigate these reconfigurations of agency, trust, and power. Doing so not only highlights opportunities for patient empowerment but also surfaces the profound ethical tensions that arise from it. This raises critical questions around how to design for \textbf{responsible empowerment}: How can we support patient agency while mitigating the risks of misinformation? What mechanisms are needed for accountability when this fragile agency leads to harm? By foregrounding these dynamics, our study emphasizes the need to carefully design and evaluate patient-LLM–provider relationships in real-world healthcare settings. It makes clear that LLMs do not simply act as assistants but actively and fundamentally reshape the socio-technical ecosystem of care, with all the opportunities and perils that they entail.

\subsection{\yuling{Situating LLM in the Chinese Healthcare System}}
\label{sub:discussion:china}

\yuling{
Our study was conducted within the Chinese healthcare system, characterized by hospital-centric care, high patient volumes, and a still-developing primary care infrastructure~\cite{chen2021ten, liang2021patient}. Particularly, the highly strained medical resources, limited consultation time, and strong culture of professional authority often mean that physicians in China hold absolute epistemic dominance, while patients are generally expected to assume a compliant and passive role in accepting diagnoses and treatment plans \cite{liang2021patient, balint1955doctor}. Within this context, patients typically lack the space for two-way communication with their doctors and are deprived of channels to both obtain information and express their preferences.
Our study, however, reveals that the introduction of LLMs is fundamentally restructuring the long-standing knowledge asymmetry–power inequality paradigm.
}

\review{
First, the LLM functioned as a proxy for missing primary care, filling a specific structural void. Unlike systems with established gatekeepers (\eg primary care providers, PCPs), Chinese patients often bear the burden of self-triage~\cite{yin2022hypertension}. In our study, the LLM helped patients identify critical signals and filter relatively minor illnesses, which are typically performed by PCPs or general practitioners (GPs) in other systems. By stitching together fragmented episodes of care, the LLM acted less as a substitute for specialists and more as essential infrastructure for the missing generalist layer.
Second, the pervasive phenomenon of long waiting times and short consultation times~\cite{yan2021patient} positioned the LLM as an \textit{overflow valve} for clinical pressure. The extreme time scarcity in public hospitals often forces a transactional efficiency where emotional support and detailed education are sacrificed. The LLM provided a psychologically safer space for asking naive questions and exploring alternatives without the pressure of clinical hierarchies. This allowed patients to effectively extend the clinical encounter, obtaining the detailed sensemaking (Section~\ref{finding4.1}) and emotional buffering (Section~\ref{finding3.1}) that the current human-based system could not provide.
Third, China’s rapid digitization and the proliferation of domestic LLMs (e.g., Doubao, DeepSeek) facilitated a rapid normalization of these roles. The low barrier to adoption and high cultural familiarity with mobile health technologies likely accelerated the depth of integration we observed, enabling patients to quickly treat LLMs as companions rather than just tools.
}

\yuling{It is important to note that patients' use of LLMs is not inherent, but rather shaped by the surrounding healthcare and cultural context \cite{suchman2007human, star1994steps}. The specific structural and cultural characteristics of Chinese healthcare system not only influence how patients incorporate LLMs into their practices, but also amplify the potential impact LLMs can have within this environment. In such a setting, LLMs do not merely function as instruments for information provision or decision support, but also as a new socio-technical mediator, filling gaps left by an underdeveloped primary-care system, extending the temporal and emotional boundaries of clinical encounters \cite{yan2021patient, guadagnoli1998patient}, and reconfiguring long-standing asymmetries in medical knowledge and power. The effectiveness of LLMs thus lies not only in their technical performance, but in how they become embedded across patients’ longitudinal healthcare journeys, taking on roles that were previously distributed across disparate institutions, personnel, and systems.}

\yuling{
More importantly, our study implies that the greatest impact of LLMs may not emerge in resource-rich, structurally mature healthcare systems, but rather in settings where resources are strained, knowledge asymmetries are pronounced, and patients shoulder substantial burdens like China. In such environments, LLMs shift from being optional add-on tools to becoming adaptive companions and emergent infrastructures woven throughout patients' everyday healthcare practices. 
Building on our findings, we call for a reconsideration of how LLMs should be evaluated and designed for healthcare. LLMs' value cannot be adequately assessed solely through diagnostic accuracy or performance on isolated tasks; instead, it must be understood within the real, continuous, relational dynamics of healthcare-seeking. By foregrounding these dynamics, we emphasize the need for future design to navigate the complex coexistence of empowerment and risk, and to carefully shape new forms of collaboration among patients, LLMs, and healthcare providers.}

\subsection{Risks in the Dark when Extending LLMs beyond Decision-Support Tools}
\label{sub:discussion:risks}
Despite LLMs' known risks of information inaccuracy, hallucinations, and lack of contextual understanding that we mentioned in Section~\ref{sub:related_work:2.2}, as well as the potential negative emotion impact in Section~\ref{finding3.3}, \review{people have already started to embrace them in their healthcare-seeking journeys beyond simple Q\&A tools~\cite{xiao2024understanding,song2025typing,jung2025ve}.}
In addition to the fragile agency mentioned above, there are other safety and ethical risks that emerge when patients start to ascribe roles to LLMs that go far beyond decision-support. 
These risks are often subtle, operating ``in the dark'' because they are intertwined with the very benefits we observed.
 
For instance, the trade-off is most apparent in the LLM's role as an emotional companion. LLMs could provide everyday reassurance and an ``emotional buffer'' in critical moments, alleviating patient anxiety (Sections~\ref{finding3-emotional}). The risk, however, is the potential for misplaced calm. A patient might be reassured by an LLM's gentle and comprehensive-sounding explanation for a symptom that, in reality, warrants urgent medical attention. Here, the tool's effectiveness in providing emotional comfort could directly lead to a delay in care.
Furthermore, a long-term reliance on a non-human agent for emotional support in health matters could create an unhealthy dependency, masking underlying mental health issues like severe health anxiety that require professional human intervention. The companion that makes a patient feel "respected and understood"  might also inadvertently isolate them from human support systems.
\minorreview{This mirrors the risks of ``artificial intimacy'' identified recently in the HCI community~\cite{skjuve2021my, brandtzaeg2022my}, where the simulation of empathy fosters an ``illusion of companionship''~\cite{arnd2015sherry, chu2025illusions}.} \review{In high-stakes health contexts, this illusion is particularly risky, as it may obscure the system's lack of genuine accountability or capability for crisis intervention~\cite{yoo2025ai}.
}

A similar tension exists in the LLM's informational and cognitive scaffolding roles. Patients felt empowered by LLMs that helped them prepare for clinical encounters and translate complex medical advice into actionable steps (Sections~\ref{finding2-informational} and \ref{finding4-cognitive}).
\review{
Personal informatics literature has long cautioned that digital tools do not simply reflect reality but actively curate it, often rendering specific aspects of illness invisible to clinicians~\cite{kirchner2021they, chung2016boundary}.
}
When an LLM helps a patient prepare to describe their symptoms, it is not merely organizing information; it is shaping a narrative. If the LLM's underlying analysis is flawed, it could prime the patient to emphasize irrelevant details while omitting critical ones, sending the human clinician down the wrong diagnostic path.
Likewise, when the LLM ``decompressed'' a doctor's concise advice into detailed daily actions, it filled the gaps with its own logic. An error here is not a simple factual mistake but a plausible-sounding, actionable instruction that could be subtly incorrect and harmful over time, a risk the patient has little capacity to evaluate.
\review{
Notably, such risks appear more pronounced as AI literacy decreases. As participants with higher literacy tended to draw on prior experience, enabling them to remain attentive upon potential errors and maintain calibrated judgments about the system’s reliability. Conversely, lower-literacy users often lacked the capability understanding required to judge response credibility. While this reduced sensitivity to errors potentially increased users' satisfaction with the tool, it also left them more susceptible to over-trust and its potential harms.
This vulnerability echoes the automation bias~\cite {cummings2017automation}, exacerbating the risk of users accepting harmful advice simply because it is articulated with professional tone and structure.
}

\review{Data privacy concerns constitute another critical dimension of the risks. LLMs themselves may inadvertently leak user-uploaded private information through various mechanisms~\cite{wang2023decodingtrust, kim2023propile}, such as potential misuse~\cite{kumar2024ethics} or incorporation into future model training~\cite{zanella2020analyzing}. The increasingly powerful capabilities of LLMs encourage users to share rich, in-depth, and even private content~\cite{li2024human, ali2025understanding}. Furthermore, the illusion of intimacy mentioned above may lead users to conflate human-like empathy with human-like privacy accountability~\cite{kwesi2025exploring, zhang2024s}.
This false sense of security leads to inadvertent sensitive disclosure, which is significantly amplified in the healthcare domain.
We also observed similar phenomena in our study, where a lack of insight into LLM mechanisms fosters misplaced trust, particularly among users with relatively lower AI literacy.
Addressing this challenge demands a multifaceted strategy such as local data anonymization~\cite{kan2023protecting, chen2023hide, wiest2025deidentifying}, enhanced patient privacy literacy~\cite{liu2025prevalence}, and robust regulatory frameworks~\cite{rathod2025privacy}.
}

Ultimately, as LLMs become longitudinal boundary companions, we must be concerned with more than just the validity of their individual outputs.
The greater risk may lie in their cumulative, often invisible, influence on the patient's entire healthcare journey.
\review{The advantages that make the LLM a powerful companion~\cite{hollanek2025ai,zhang2025rise}—its ability to traverse cognitive, emotional, and behavioral domains—also make its potential for harm more systemic and harder to pinpoint.}
In addition to correcting factual errors from LLMs, future LLM-based health applications should also be fully aware of and prepared for the significant risks of the effects of a subtly skewed communication dynamic, a falsely reassured emotional state, or a pattern of self-care built on a foundation of flawed sensemaking.
These are the risks that lurk in the dark, and they demand that our focus shift from merely ensuring accuracy to safeguarding the integrity of the patient's holistic experience and decision-making process.

\subsection{Design Implications for Future LLM-Powered Healthcare Applications}
\label{sub:discussion:design}

Our findings on the multifaceted roles of LLMs offer critical insights for the future design of LLMs in healthcare. We propose the following design directions that embrace the LLM's potential as a boundary companion while mitigating the risks associated with reconfigured patient agency and the subtle dangers that emerge.

\textbf{Designing for Relational Continuity, Not Just Contextual Awareness.}
Our study shows that patients want to engage with LLMs as longitudinal companions, yet current models with short-term memory often fail to support this, leading to frustration (Section \ref{finding3.3}). 
\minorreview{Long-term memory for chatbots is an active research direction, which could mitigate this issue by fostering familiarity, which could encourage users to disclose health information, strengthen their perception of the system, and form long-term relationships or even ~\cite{jo2024understanding}. }
\review{However, the design challenge is not merely to add memory but to design for relational continuity, a concept supported by relational agent literature, which suggests that maintaining a sense of shared history and "remembering" past interactions is fundamental to building alliances~\cite{bickmore2005establishing, bickmore2010maintaining}.}
Therefore, instead of one-off Q\&A sessions, future systems should be designed to build and maintain a relationship over a patient's entire healthcare journey. This could involve: 
(i) Dynamic patient summaries: Creating evolving summaries that capture not only clinical facts (\eg diagnoses, medications) but also patient-specific context, such as their communication preferences, emotional state, and key concerns expressed over time;
and (ii) Integrating patient-provider dialogue: With patient consent, the LLM could be aware of key advice from clinicians. This would enable it to act as a more effective translator and scaffolder, helping patients turn "the clinician's advice into specific daily actions" without contradicting the core treatment plan. \review{This reinforces the LLM's complementary role and helps maintain a cohesive care narrative~\cite{chung2016boundary,kirchner2021they}.}
Crucially, grounding the LLM's continuous support in the clinician's expert guidance serves as a critical check against the systemic risk of the LLM subtly misdirecting a patient's journey over time.

\textbf{Designing for Responsible Empowerment and Trust Calibration.}
Our concept of fragile agency highlights a critical ethical tension: LLMs empower patients but also expose them to risks from misinformation, as they may lack the expertise to vet the LLM's advice. The design goal must shift from simply warning users about inaccuracies to actively designing for responsible empowerment. 
This is especially vital given the risks are not just overt error but also misplaced calm that can delay care or subtle misdirection that can flaw communication with clinicians.
\review{Existing work on human-AI collaboration suggests that explicitly signaling system limitations or uncertainty can effectively calibrate user trust and reduce automation bias~\cite{kocielnik2019will, wang2025impact,okamura2020adaptive}.}
This requires moving beyond static disclaimers and creating interfaces that help patients calibrate their trust in real time.
For instance, instead of delivering all information with the same confident tone, \review{LLMs could be designed to express uncertainty explicitly~\cite{tomsett2020rapid}}. When discussing less-established treatments or ambiguous symptoms, the UI could visually differentiate speculative information from evidence-based guidelines, with phrases such as ``Here are a few possibilities being explored by researchers, but they are not standard practice...'', \review{acting as a direct countermeasure to the risk of false reassurance with overly confident tone~\cite{okamura2020adaptive}.}
Besides, rather than being an answer machine, the LLM could act as a Socratic partner. When a patient considers a significant action (\eg challenging a doctor's plan), the LLM could prompt critical reflection: ``What are the main reasons you're considering this alternative? What are your biggest concerns about the doctor's original recommendation?'' This approach empowers patients not by giving them answers, but by equipping them with better questions and a more structured thought process, \review{thereby making their agency more robust~\cite{usmani2023human,kang2022ai}.}

\textbf{From Information-Giving to Scaffolding Health Literacy.}
Our findings show that one of the LLM's cognitive roles was to decompress dense medical advice into understandable, actionable knowledge (Section \ref{finding4.1}). \review{Future designs should formalize this educational role by focusing on scaffolding long-term health literacy, as emphasized by the personal informatics community on going beyond providing data but supporting users' sensemaking~\cite{puussaar2017enhancing, mamykina2015adopting}.}
To ensure genuine understanding, the LLM could employ a "teach-back" method, a proven clinical communication technique~\cite{tran2019teach,shersher2021definitions}. After explaining a concept, it could ask the patient to tell the LLM back in the patient's own words what this means. This transforms passive information consumption into active learning.
Such interactive validation is a crucial safeguard, ensuring the patient has truly understood the nuances of the advice, and mitigating the risk of them acting on subtly misinterpreted or incorrectly "decompressed" guidance.
Meanwhile, LLM systems could move from providing lists of suggestions to collaboratively creating personalized plans with the patient. For example, after discussing post-operative recovery, the LLM could help the patient co-author a simple daily schedule that incorporates the doctor's advice on medication, diet, and light exercise, making abstract guidance concrete and integrated into their lived reality. In this way, the LLM serves as a sustained partner, building the patient's capacity and confidence to navigate their own health journeys.

\textbf{Self-Guardian Infrastructures to Improve Reliability.}
In addition to the patient-facing aspects of design, adopting a self-guardian and self-tracking infrastructure is a potential solution to reduce risks from the technical perspective. While the above suggestions focus on shaping the human-AI interaction, the internal architecture of the LLM system itself can be re-envisioned to be inherently safer and more reliable, \review{reflecting the broader ``Human-Centered AI'' framework, which advocates for reliable, auditable, and safe system architectures as a prerequisite for user trust~\cite{shneiderman2020human, amershi2019guidelines}.}
A multi-agent system is a viable solution for this~\cite{heydari2025anatomy,kim2024mdagents}, moving beyond a single LLM to a collaborative ensemble of specialized agents that check and balance one another, where distinct functions would be responsible for oversight.
Some capabilities of such an infrastructure would include: continuous internal validation (systematically cross-referencing generated content against authenticated medical knowledge bases or established clinical guidelines), and dynamic safety monitoring (assessing conversational context to identify potential risks).
By embedding these self-monitoring and corrective functions directly into the system's architecture, the burden of validation is shifted away from the patient. This approach provides a more robust technical foundation for responsible empowerment, allowing the LLM to more safely fulfill its complex roles as a longitudinal healthcare companion.

\subsection{Limitations and Future Work}
\label{sub:discussion:limitations}
Our study has several limitations that point to important directions for future work. First, our qualitative analysis is based on patients' diary reflections. While the data is rich, they do not include a fine-grained analysis of the verbatim patient-LLM conversation data that could reveal deeper interactional nuances.
Second, as discussed above, our study was conducted with 25 patients in China using specific LLMs available at the time; findings may differ across other cultural contexts, healthcare systems, and with more advanced or specialized LLM technologies.
\review{
We posit that our core findings are likely generalizable to other healthcare systems facing similar pressures of time-scarcity and provider burnout globally~\cite{writing2022report,tai2007time},
yet future cross-cultural research is needed to disentangle these universal technological affordances from culturally specific adoption patterns.
}
Meanwhile, participants who joined the study may inherently favor the adoption of LLMs, which may lead to potential bias in our data analysis.
\review{Additionally, while our study spanned four weeks (which is already longer or comparable to other similar studies~\cite{hong2020using,fu2024text}), future longer-term research could uncover the evolution over an extended timeframe of patients' sense of connection with the LLM, perceived advice helpfulness, and reliance patterns.
}
Furthermore, our findings are centered entirely on the patient's perspective, intentionally bracketing the crucial viewpoint of clinicians. Future research should therefore analyze these dynamics from multiple perspectives, including those of healthcare providers, to develop a holistic understanding of the reconfigured patient–doctor–AI triad.

\section{Conclusion}
\label{sec:conclusion}

In this paper, we presented a longitudinal diary study to examine how patients integrate LLMs into their healthcare-seeking journeys. Our findings demonstrated that patients adopt LLMs not as simple diagnostic or decision-support tools, but as dynamic longitudinal boundary companions that scaffold their experiences across informational, behavioral, emotional, and cognitive levels. 
This integration empowers patients as more active negotiators of their own care by reshaping relationships of agency, trust, and power.
Our work suggests future directions for creating socio-technical systems that support relational continuity, foster responsible empowerment, and contribute to a safer and collaborative healthcare ecosystem across patients, LLMs, and clinicians.

\begin{acks}
\minorreview{Research reported in this publication was supported in part by the National Science Foundation (NSF) SCH under Award Number 2306690. The content is solely the responsibility of the authors and does not necessarily represent the official views of the National Science Foundation. We also thank the Columbia Research Stabilization Grant for supporting this research project. We want to thank all participants for taking part in our study.} We acknowledge the use of Generative AI tools to improve the grammar, style, and readability of this manuscript. These tools were used exclusively for text editing and played no role in the qualitative data analysis, interpretation, or generation of the core findings presented.
\end{acks}


\newpage
\onecolumn  
\section*{Appendix: }
\label{appendix:}

\begin{table*}[h]
\centering
\caption{Details of all LLMs used by participants in this study, categorized as general-purpose and health-focused.}
\small
\renewcommand{\arraystretch}{1.2}
\begin{tabular}{ c l l l }
\toprule
\textbf{LLM Category} & \textbf{Specific LLM Name} & \textbf{Developer} & \textbf{URL} \\
\midrule
\multirow{5}{*}{\parbox{2cm}{\centering General-purpose LLM}} & Doubao & ByteDance & \url{https://www.doubao.com/chat/} \\
& DeepSeek & DeepSeek AI & \url{https://www.deepseek.com/} \\
& Qwen & Alibaba Cloud & \url{https://www.tongyi.com/} \\
& Kimi & Moonshot AI & \url{https://www.kimi.com/} \\
& ChatGPT & OpenAI & \url{https://chatgpt.com/} \\
\midrule
\multirow{6}{*}{\parbox{2cm}{\centering Health-focused LLM}} & Xiaohe Health & ByteDance & \url{https://www.tiktok.com/explore} \\
& iFlytek Xiaoyi & iFlytek & \url{https://chatdr.iflyhealth.com/} \\
& Baidu Lingyi Zhihui & Baidu & \url{https://01.baidu.com/home/} \\
& JD Health & JD Health & \url{https://ir.jdhealth.com/sc/index.php} \\
& Zuoshouyisheng & Zuoshouyisheng & \url{https://zuoshouyisheng.com/} \\
& AQ & Ant Group & \url{https://www.alipay.com/} \\
\bottomrule
\end{tabular}
\label{tab:LLM_Tools}
\end{table*}

\begin{table*}[h!]
\centering
\caption{Details of all diary questionnaires, including interaction contexts, the corresponding question categories, and the specific questions.}
\label{tab:DiaryQuestionnaire}
\small
\renewcommand{\arraystretch}{1.3}
\resizebox{\textwidth}{!}{%
\begin{tabular}{ m{2.8cm} L{3.2cm} L{11cm} }
\toprule
\textbf{Interaction Context} & \textbf{Question Categories} & \textbf{Specific Question} \\
\midrule
\multirow[c]{9}[10]{2.75cm}{\raggedright\arraybackslash Interaction with LLM(s)}
    & \multirow{2}{*}{Specifics}      & Did you consult more than one AI tool for health advice today? \\
    &                                 & Which LLM(s) did you use? \\
\cmidrule(lr){2-3}
    & \multirow{2}{=}{Rationale for the Interaction} & What was your primary reason for seeking health advice from the LLM today? \\
    &                                 & [If applicable] What were your reasons for consulting multiple LLMs?  \\
\cmidrule(lr){2-3}
    & Conversation Content            & Please provide the transcript (via link or screenshot) of your LLM conversation(s) today.  \\
\cmidrule(lr){2-3}
    & Perceived Role                  & What role did the LLM play in your health trajectory? \\
\cmidrule(lr){2-3}
    & \multirow{3}[2]{*}{User Experience} & Please describe your overall experience with the LLM interaction today.\\
    &                                 & How did the LLM influence your emotional state today?\\
    &                                 & Do you have any feedback on your interaction with the LLM today, such as unsatisfactory aspects or suggestions for future improvements?\\
\midrule
\multirow[c]{3}[4]{2.75cm}{\raggedright Interaction with human doctor(s)}
    & Reason for the Visit            & What was your primary reason for consulting human doctor(s) today?  \\
\cmidrule(lr){2-3}
    & Clinical Outcomes               & Please summarize what you have discussed with the human doctor today.  \\
\cmidrule(lr){2-3}
    & Perceived Role                  & What role did the human doctor's advice play in your health trajectory? \\
\midrule
\multirow[c]{3}[8]{2.75cm}{\raggedright Interaction with both}
    & Reason for Interacting with Both & What were your reasons for consulting both the LLM and the human doctor for your health concern today?  \\
\cmidrule(lr){2-3}
    & \multirow{2}[2]{=}{Perceived Relationship between the Two} & What were the key differences between the advice provided by the LLM and the human doctor?  \\
    &                                 & How do you perceive the relationship between the LLM and the human doctor in the context of the healthcare-seeking trajectory? \\
\midrule
Others & - & Do you have any other things you would like to share? \\
\bottomrule
\end{tabular}%
}
\end{table*}

\begin{table*}[t]
\centering
\caption{Participants' educational backgrounds and AI literacy levels. We report four representative items from the AI literacy scale, alongside the Total Score, which is calculated as the mean of all \review{12 items \cite{wang2023measuring}}. All scores in this table range from 1 to 7, with higher values indicating higher AI literacy.}
\renewcommand{\arraystretch}{1.2}
\resizebox{\textwidth}{!}{%
\begin{tabular}{ 
    >{\centering\arraybackslash}p{1.0cm}   %
    >{\centering\arraybackslash}p{1.8cm}   %
    >{\centering\arraybackslash}p{2.2cm}   %
    >{\centering\arraybackslash}p{2.2cm}   %
    >{\centering\arraybackslash}p{2.2cm}   %
    >{\centering\arraybackslash}p{2.2cm}   %
    >{\centering\arraybackslash}p{2.2cm}   %
    }
\toprule
\raisebox{-1.5ex}{\textbf{ID}} & \raisebox{-1.5ex}{\textbf{Education}} & \textbf{Device Distinction} & \textbf{Tech Identification} & \textbf{Capability Evaluation} & \textbf{Misuse Vigilance} & {\textbf{Total Score (averaged)}} \\
\midrule
P01 & Bachelor's & \progressratio[7]{7}{1}{7} & \progressratio[7]{5}{1}{7} & \progressratio[7]{5}{1}{7} & \progressratio[7]{5}{1}{7} & \progressratioTotal[7]{5.8}{1}{7} \\
P02 & Bachelor's & \progressratio[7]{6}{1}{7} & \progressratio[7]{6}{1}{7} & \progressratio[7]{5}{1}{7} & \progressratio[7]{5}{1}{7} & \progressratioTotal[7]{6.1}{1}{7} \\
P03 & Bachelor's & \progressratio[7]{7}{1}{7} & \progressratio[7]{6}{1}{7} & \progressratio[7]{7}{1}{7} & \progressratio[7]{5}{1}{7} & \progressratioTotal[7]{6.2}{1}{7} \\
P04 & Bachelor's & \progressratio[7]{7}{1}{7} & \progressratio[7]{5}{1}{7} & \progressratio[7]{5}{1}{7} & \progressratio[7]6{1}{7} & \progressratioTotal[7]{6.0}{1}{7} \\
P05 & Bachelor's & \progressratio[7]{6}{1}{7} & \progressratio[7]{5}{1}{7} & \progressratio[7]{5}{1}{7} & \progressratio[7]{4}{1}{7} & \progressratioTotal[7]{5.4}{1}{7} \\
P06 & Bachelor's & \progressratio[7]{6}{1}{7} & \progressratio[7]{5}{1}{7} & \progressratio[7]{6}{1}{7} & \progressratio[7]{5}{1}{7} & \progressratioTotal[7]{6.3}{1}{7} \\
P07 & Bachelor's & \progressratio[7]{6}{1}{7} &  \progressratio[7]{5}{1}{7} & \progressratio[7]{3}{1}{7} & \progressratio[7]{2}{1}{7} & \progressratioTotal[7]{5.3}{1}{7} \\
P08 & Bachelor's & \progressratio[7]{5}{1}{7} & \progressratio[7]{5}{1}{7} & \progressratio[7]{5}{1}{7} & \progressratio[7]{3}{1}{7} & \progressratioTotal[7]{4.8}{1}{7} \\
P09 & Master's & \progressratio[7]{7}{1}{7} & \progressratio[7]{7}{1}{7} & \progressratio[7]{7}{1}{7} & \progressratio[7]{7}{1}{7} & \progressratioTotal[7]{6.6}{1}{7} \\
P10 & Bachelor's & \progressratio[7]{5}{1}{7} & \progressratio[7]{6}{1}{7} & \progressratio[7]{6}{1}{7} & \progressratio[7]{5}{1}{7} & \progressratioTotal[7]{5.9}{1}{7} \\
P11 & Bachelor's & \progressratio[7]{7}{1}{7} & \progressratio[7]{6}{1}{7} & \progressratio[7]{6}{1}{7} & \progressratio[7]{4}{1}{7} & \progressratioTotal[7]{6.5}{1}{7} \\
P12 & Bachelor's & \progressratio[7]{6}{1}{7} & \progressratio[7]{7}{1}{7} & \progressratio[7]{6}{1}{7} & \progressratio[7]{3}{1}{7} & \progressratioTotal[7]{5.1}{1}{7} \\
P13 & Bachelor's & \progressratio[7]{6}{1}{7} & \progressratio[7]{5}{1}{7} & \progressratio[7]{6}{1}{7} & \progressratio[7]{5}{1}{7} & \progressratioTotal[7]{6.0}{1}{7} \\
P14 & Bachelor's & \progressratio[7]{7}{1}{7} & \progressratio[7]{7}{1}{7} & \progressratio[7]{7}{1}{7} & \progressratio[7]{1}{1}{7} & \progressratioTotal[7]{6.5}{1}{7} \\
P15 & Bachelor's & \progressratio[7]{6}{1}{7} & \progressratio[7]{7}{1}{7} & \progressratio[7]{7}{1}{7} & \progressratio[7]{2}{1}{7} & \progressratioTotal[7]{6.0}{1}{7} \\
P16 & Bachelor's & \progressratio[7]{7}{1}{7} & \progressratio[7]{5}{1}{7} & \progressratio[7]{5}{1}{7} & \progressratio[7]{1}{1}{7} & \progressratioTotal[7]{5.3}{1}{7} \\
P17 & Bachelor's & \progressratio[7]{7}{1}{7} & \progressratio[7]{5}{1}{7} & \progressratio[7]{4}{1}{7} & \progressratio[7]{5}{1}{7} & \progressratioTotal[7]{6.0}{1}{7} \\
P18 & Bachelor's & \progressratio[7]{7}{1}{7} & \progressratio[7]{6}{1}{7} & \progressratio[7]{5}{1}{7} & \progressratio[7]{5}{1}{7} & \progressratioTotal[7]{5.8}{1}{7} \\
P19 & Bachelor's & \progressratio[7]{7}{1}{7} & \progressratio[7]{5}{1}{7} & \progressratio[7]{5}{1}{7} & \progressratio[7]{7}{1}{7} & \progressratioTotal[7]{6.5}{1}{7} \\
P20 & Bachelor's & \progressratio[7]{6}{1}{7} & \progressratio[7]{6}{1}{7} & \progressratio[7]{5}{1}{7} & \progressratio[7]{5}{1}{7} & \progressratioTotal[7]{5.4}{1}{7} \\
P21 & Bachelor's & \progressratio[7]{7}{1}{7} & \progressratio[7]{3}{1}{7} & \progressratio[7]{7}{1}{7} & \progressratio[7]{4}{1}{7} & \progressratioTotal[7]{5.8}{1}{7} \\
P22 & Bachelor's & \progressratio[7]{7}{1}{7} & \progressratio[7]{7}{1}{7} & \progressratio[7]{7}{1}{7} & \progressratio[7]{2}{1}{7} & \progressratioTotal[7]{5.9}{1}{7} \\
P23 & Bachelor's & \progressratio[7]{6}{1}{7} & \progressratio[7]{6}{1}{7} & \progressratio[7]{4}{1}{7} & \progressratio[7]{5}{1}{7} & \progressratioTotal[7]{5.7}{1}{7} \\
P24 & High School & \progressratio[7]{4}{1}{7} & \progressratio[7]{5}{1}{7} & \progressratio[7]{2}{1}{7} & \progressratio[7]{1}{1}{7} & \progressratioTotal[7]{5.3}{1}{7} \\
P25 & Bachelor's & \progressratio[7]{5}{1}{7} & \progressratio[7]{6}{1}{7} & \progressratio[7]{6}{1}{7} & \progressratio[7]{6}{1}{7} & \progressratioTotal[7]{5.9}{1}{7} \\
\bottomrule
\end{tabular}%
}
\label{tab:AI_Literacy}
\end{table*}

\end{CJK*}
\end{document}